\documentclass[aps, prd, onecolumn]{revtex4-2}

\usepackage{subfig}
\usepackage{amsmath}
\usepackage{amssymb}
\usepackage{amsthm}
\usepackage{mathrsfs}
\usepackage{graphicx}
\usepackage{fancyhdr}
\usepackage{array}
\usepackage{geometry}
\usepackage{bm}
\usepackage{latexsym}
\usepackage[all]{xy}
\usepackage{euscript}
\usepackage{tensor}
\usepackage{enumerate}
\usepackage{slashed}
\usepackage{accents}
\usepackage{comment}
\usepackage{mathrsfs}
\usepackage{hyperref}
\usepackage{tensor}
\usepackage{ulem}
\usepackage{xcolor}
\usepackage{cancel}
\usepackage{color}
\usepackage{setspace}

\newcommand{\Christ}[3]{\ensuremath{\left\{ {}\indices{_#2}\!{}\indices{^#1}{}\!\indices{_#3} \right\}}}

\makeatletter
\newcommand*{\leqdef}{\mathrel{\rlap{%
		\raisebox{0.25ex}{$\m@th\cdot$}}%
	\raisebox{-0.25ex}{$\m@th\cdot$}}%
=}
\newcommand*{\reqdef}{=\mathrel{\rlap{%
		\raisebox{0.25ex}{$\m@th\cdot$}}%
	\raisebox{-0.25ex}{$\m@th\cdot$}}
}

\newcommand*{\leqdefchart}{\mathrel{\rlap{%
		\raisebox{0.25ex}{$\m@th\cdot$}}%
	\raisebox{-0.25ex}{$\m@th\cdot$}}%
\overset{*}{=}}
\newcommand*{\reqdefchart}{\overset{*}{=}\mathrel{\rlap{%
		\raisebox{0.25ex}{$\m@th\cdot$}}%
	\raisebox{-0.25ex}{$\m@th\cdot$}}
}

\newcommand*{\hatnabla}{\widehat{\nabla}}

\newcommand*{\multifield}[4]{\underset{#2}{#1}\tensor{}{^{#3}_{#4}}}
\newcommand*{\multifieldamp}[4]{\underset{#2}{#1}{}^{#3}_{#4}}

\newcommand\munderbar[1]{\underaccent{\bar}{#1}}
\def\nunderbar#1{\underline{\sbox\tw@{$#1$}\dp\tw@\z@\box\tw@}}

\newcommand{\bigp@rp}[2]{%
\vcenter{
	\m@th\hbox{\scalebox{\ifx#1\displaystyle1.5\else1.5\fi}{$#1\perp$}}
}%
}

\newcommand{\bigperp}{%
\mathop{\mathpalette\bigp@rp\relax}%
\displaylimits
}

\usepackage{tikz}
\usetikzlibrary{shapes.geometric, arrows}
\usetikzlibrary{fit}

\makeatother

\begin{document}

\title{A unified approach to coupled homogeneous \\ linear wave propagation in generic gravity}
	
\author{Lucas T. Santana}
\email{lts@if.ufrj.br}
\affiliation{Universidade Federal do Rio de Janeiro,
		Instituto de F\'\i sica, \\
		CEP 21941-972 Rio de Janeiro, RJ, Brazil}

\author{Jo\~ao C. Lobato}
\email{jcavlobato@if.ufrj.br}
\affiliation{Universidade Federal do Rio de Janeiro,
		Instituto de F\'\i sica, \\
		CEP 21941-972 Rio de Janeiro, RJ, Brazil}

\author{Ribamar R. R. Reis}
\email{ribamar@if.ufrj.br}
\affiliation{Universidade Federal do Rio de Janeiro,
		Instituto de F\'\i sica, \\
		CEP 21941-972 Rio de Janeiro, RJ, Brazil}
\affiliation{Universidade Federal do Rio de Janeiro, Observat\'orio do Valongo, 
		\\CEP 20080-090 Rio de Janeiro, RJ, Brazil}
	
\author{Maur\'\i cio O. Calv\~ao}
\email{orca@if.ufrj.br}
\affiliation{Universidade Federal do Rio de Janeiro,
		Instituto de F\'\i sica, \\
		CEP 21941-972 Rio de Janeiro, RJ, Brazil}

\begin{abstract}
    Wave propagation is a common occurrence in all of physics. A linear approximation provides a simpler way to describe various fields related to observable phenomena in laboratory physics as well as astronomy and cosmology, allowing us to probe gravitation through its effect on the trajectories of particles associated with those fields. This paper proposes a unified framework to describe the wave propagation of a set of interacting tensor fields that obey coupled homogeneous linear second-order partial differential equations for arbitrary curved spacetimes, both Lorentzian and metric-affine. We use JWKB Ans\"atze for all fields, written in terms of a perturbation parameter proportional to a representative wavelength among them, deriving a set of hierarchical algebraic and differential equations that link the fields' phases and different order amplitudes. This allows us to reobtain the well-known laws of geometrical optics and beyond geometrical optics in a generalized form, showing that these laws are independent of the rank of the fields involved. This is true as long as what we refer to as the kinetic tensor of a given field satisfies a set of diagonality conditions, which further imply a handful of simplifications on the transport equations obtained in the subleading orders of the JWKB Ans\"atze. We explore these results in several notable examples in Lorentzian and metric-affine spacetimes, illustrating the reach of our derivations in general relativity, reduced Horndeski theories, spacetimes with completely antisymmetric torsion and Weyl spacetimes. The formalism presented herein lays the groundwork for the study of rays associated with different types of waves in curved spacetimes and provides the tools to compute modifications to their brightness evolution laws, consequential distance duality relations, and beyond geometrical optics phenomena.
\end{abstract}

\maketitle

\section{Introduction}
\label{sec:introduction}

Describing the propagation of linear waves in curved spacetimes is of uttermost importance to model several physical phenomena of relevance to modern cosmology and astrophysics, such as gravitational lensing \cite{Schneider1992} and the propagation of gravitational waves \cite{Abbott2016}. But describing (scalar, vector, tensor, etc.) linear waves associated to a physical field directly by means of the corresponding equations of motion in curved spacetimes is generally a difficult endeavor, and different workarounds are usually employed to tackle this issue in a simpler or more tractable form. For one, there is the study of discontinuities in a field by means of so-called \textit{characteristic surfaces} and \textit{bi-characteristic curves} in general \cite{Hadamard1923, Bremmer1951, Kline1965, Friedlander1975, Courant1989, Born1999, Bona2011} as well as in the case, \textit{e.g.}, of modified gravity \cite{Reall2014, Tanahashi2017}. Additionally, \textit{Fourier transforms} allow us to study the oscillation modes of a physical field (\textit{e.g.} the electromagnetic field) \cite{Born1999, Horsley2011}, although such a decomposition might only be possible in specific spacetimes possessing a number of symmetries. 

Of particular relevance is an alternative procedure given by an approximation scheme where a field is associated to a phase function and a family of amplitudes, combined in a (formal) Jeffreys-Wentzel-Kramers-Brillouin (JWKB) expansion [cf. Eqs.~(\ref{eqs:full-rewritten-ansatze})] \cite{Ehlers1967, Misner1973, Anile1989, Schneider1992, Perlick2000, ChoquetBruhat2009, Ellis2012}. This relies on well-known observables reminiscent from general physics, such as wavelength, intensity, and polarization. In particular, the wavelength of a wave---inherently associated to a reference frame---introduces a characteristic scale allowing for two crude regimes of wave propagation to be identified: (i) one in which the wavelength is much smaller than any other length scale in the region where the wave is passing by, and (ii) another where the wavelength is of the order of (or bigger than) some (or all) length scales in the considered region. The former assumes that the wave character of the field might essentialy be disregarded, and a family of rays is identified as the trajectories traversed by corresponding fiducial particles. These rays are geometrical paths in spacetime, which make this regime generally referred to as \textit{geometrical optics} (GO) or \textit{eikonal} or \textit{high frequency approximation} [cf. Eqs.~(\ref{eq:leading-order-general}) and (\ref{eq:1st-subleading-order-general})], a nomenclature common to electromagnetic waves and also used when referring to scalar fields and gravitational waves \cite{Harte2019, Cusin2019}. In contrast, the second regime, in which the wave character of a field should be more thoroughly taken into account, is usually referred to as \textit{beyond geometrical optics} (BGO) or \textit{diffraction regime} [cf. Eqs.~(\ref{eqs:hierarchy-general})] \cite{Kravtsov1990, Harte2019, Cusin2020}. Whilst BGO stands for a regime where interference is described by the nonvanishing subleading amplitudes in the JWKB approximation, the general formalism still assumes that fields satisfy systems of second-order homogeneous linear equations of motion \cite{Cusin2020}.

In this work, we propose a general framework to describe a set of tensor fields that satisfy coupled second-order linear homogeneous equations of wave-like form [\textit{viz.}, Eqs.~(\ref{eq:field-equation}) and (\ref{eq:field-equation-operator})]. We employ JWKB Ans\"atze for the fields involved and deduce several results including some ones already appearing in the known literature regarding GO and BGO in metric-affine spacetimes (which include Lorentzian spacetimes as a subcategory) \cite{Santana2017}. We show that those known results are quite universal and directly related to special forms for the kinetic tensors in the equations of motion. Particularly, the leading order results appearing therein, which lead to the first set of GO relations, are shown to be independent of what we refer to, in the equations of motion, as friction and mass tensors. Furthermore, the first subleading order result, which is also part of GO, is completely independent of the form of the mass tensors involved. On the flipside, even for simplified kinetic tensors, higher subleading order results are still generally dependent on kinetic, friction and mass tensors. We then provide several examples in Lorentzian spacetimes, and illustrate the reach of our derivations also for two distinct families of metric-affine spacetimes.

We structure this work as follows. In Sec.~\ref{sec:multi-field-derivation} we present the unified framework, introducing the equations of motion and the JWKB Ans\"atze considered to derive the general results leading to algebraic and differential constraints between the different fields' amplitudes and phase functions. In Sec.~\ref{sec:decomposition-kinetic-operator} we write a decomposition of the general kinetic tensors and explore simplified expressions for them in Subsecs.~\ref{subsec:fg-diagonal-kinetic-operator}, \ref{subsec:fc-diagonal-kinetic-operators} and \ref{subsec:fcg-diagonal-kinetic-operators}, which allows us to further interpret the results of the prior section and establish universal results mentioned above for both GO and BGO. We then use Sec.~\ref{sec:examples-lorentzian-spacetimes} to illustrate the latter results and the utility of our general framework applied to the dynamics of different rank fields in Lorentzian spacetimes, including the Klein-Gordon field, the electromagnetic field (both in its potential and Faraday tensor representations), gravitational waves in general relativity and reduced Horndeski theories (RHTs), the latter being an important example of modified scalar-tensor gravity having significant interest in the recent literature \cite{Dalang2020, Dalang2021, Lobato2022, Lobato2024}. In Sec.~\ref{sec:examples-metric-affine-spacetimes} we apply our general results to two distinct familes of metric-affine spacetimes, namely, one possessing a metric-compatible connection and a completely antisymmetric torsion, and another referred to as a Weyl spacetime. We conclude in Sec.~\ref{sec:discussion} with a discussion of our results and future perspectives. App. \ref{append:metric-affine-geometry} is dedicated to a quick review of metric-affine spacetimes.

All our derivations assume a generic metric-affine spacetime, \textit{i.e.}, a triple $(\mathscr{M}, g_{\alpha\beta}, \tensor{\Gamma}{^\mu_{\alpha\beta}})$, where $\mathscr{M}$ is a 4-dimensional smooth manifold, $g_{\alpha\beta}$, a Lorentzian metric with signature $+2$, and $\tensor{\Gamma}{^\mu_{\alpha\beta}}$ is a generic affine connection (not necessarily the Levi-Civita one). We use a wide hat over a kernel letter to denote that the corresponding geometric object is computed using the Levi-Civita connection of a given spacetime. We also choose natural units to set $c = \hbar = 1$, and whenever background fields are considered, a bar under a kernel letter is employed to denote that a quantity should be evaluated on a background field (or a set of background fields).

\section{Multifield equations of motion and JWKB Ans\"atze}
\label{sec:multi-field-derivation}

Let us start by considering an arbitrary set of tensor fields $\multifield{\Psi}{I}{A_I}{}$ in a metric-affine spacetime, where $A_I$ is a generic set of indices for the $I$-th field ($I=1, ..., M$), to be understood as a condensed notation for
\begin{equation}
	\label{eq:condensed-notation}
	\multifield{\Psi}{I}{A_I}{} \leqdef \multifield{\Psi}{I}{\alpha_1 \cdots \alpha_{r_I}}{\beta_1 \cdots \beta_{s_I}} \,,
\end{equation}
\textit{i.e.}, an element of
\begin{equation}
	\label{eq:tensor-products}
	\Pi_{\mathscr{P}}(r_I,s_I) \leqdef \left(\bigotimes_{i=1}^{r_I} T_{\mathscr{P}} \mathscr{M} \right) \otimes \left(\bigotimes_{j=1}^{s_I} T^*_{\mathscr{P}} \mathscr{M} \right)\,,	
\end{equation}
for every event $\mathscr{P} \in \mathscr{M}$. Let us then assume that these fields satisfy the following system of coupled homogeneous linear second-order partial differential equations of motion:
\begin{subequations}
\begin{equation}
	\label{eq:field-equation}
	\sum_{J=1}^M \, \multifield{\mathcal{L}}{IJ}{A_I}{B_J} \, \multifield{\Psi}{J}{B_J}{} = 0\,,
\end{equation}
where
\begin{equation}
	\label{eq:field-equation-operator}
	\multifield{\mathcal{L}}{IJ}{A_I}{B_J} \leqdef \multifield{\mathcal{K}}{IJ}{\alpha\beta A_I}{B_J} \, \nabla_\alpha \nabla_\beta + \multifield{\mathcal{F}}{IJ}{\alpha A_I}{B_J} \, \nabla_\alpha + \multifield{\mathcal{M}}{IJ}{A_I}{B_J} \,, 
\end{equation} 
\end{subequations}
such that all fields obey a \textit{linear wave} propagation (provided that the system of equations is hyperbolic), with $\nabla_\alpha$ referring to the covariant derivative with respect to the arbitrary prescribed affine connection $\tensor{\Gamma}{^\mu_{\alpha\beta}}$, not the Levi-Civita one determined solely from the metric $g_{\alpha\beta}$ [cf. App. \ref{append:metric-affine-geometry}]. $\multifield{\mathcal{K}}{IJ}{\alpha\beta A_I}{B_J} \ne 0$, $\multifield{\mathcal{F}}{IJ}{\alpha A_I}{B_J}$ and $\multifield{\mathcal{M}}{IJ}{A_I}{B_J}$ are real-valued tensor fields that do not contain zeroth or higher-order derivatives of $\multifield{\Psi}{I}{A_I}{}$ with respect to $x^\mu$, though they might depend on other prescribed tensors, such as those constructed from zeroth and/or higher-order derivatives of the metric and affine connection. We refer to them, respectively, as \textit{kinetic}, \textit{friction} and \textit{mass} tensors, following the usual terminology, which appears, for instance, in \cite{Dalang2021}. While we restrict ourselves to systems of linear equations of motion, this same nomenclature applies to systems of nonlinear equations of motion in the fields of interest, such as the full set of quasi-linear Einstein field equations in general relativity. For the sake of clarity, in all our derivations, whenever an equation includes a sum over several fields, we explicitly include a summation symbol with an uppercase Latin index ($J$ for example). Furthermore, we use a condensed version of the usual Einstein summation convention on Greek (spacetime) indices:
\begin{equation}
	\label{eq:einstein-convention-greek-indices-1}
	\multifield{\mathcal{K}}{IJ}{\alpha\beta A_I}{B_J} \, \nabla_\alpha \, \nabla_\beta \multifield{\Psi}{J}{B_J}{} \leqdef \multifield{\mathcal{K}}{IJ}{\alpha\beta\mu_1 \cdots \mu_{r_I}}{\nu_1 \cdots \nu_{s_I}}\tensor{}{_{\sigma_1 \cdots \sigma_{r_J}}^{\lambda_1 \cdots \lambda_{s_J}}} \, \nabla_\alpha \, \nabla_\beta \, \multifield{\Psi}{J}{\sigma_1 \cdots \sigma_{r_J}}{\lambda_1 \cdots \lambda_{s_J}}\,,
\end{equation}
\begin{equation}
	\label{eq:einstein-convention-greek-indices-2}
	\multifield{\mathcal{F}}{IJ}{\alpha A_I}{B_J} \, \nabla_\alpha \, \multifield{\Psi}{J}{B_J}{} \leqdef \multifield{\mathcal{F}}{IJ}{\alpha\mu_1 \cdots \mu_{r_I}}{\nu_1 \cdots \nu_{s_I}}\tensor{}{_{\sigma_1 \cdots \sigma_{r_J}}^{\lambda_1 \cdots \lambda_{s_J}}} \, \nabla_\alpha \, \multifield{\Psi}{J}{\sigma_1 \cdots \sigma_{r_J}}{\lambda_1 \cdots \lambda_{s_J}}
\end{equation}
and
\begin{equation}
	\label{eq:einstein-convention-greek-indices-3}
	\multifield{\mathcal{M}}{IJ}{A_I}{B_J} \, \multifield{\Psi}{J}{B_J}{} \leqdef \multifield{\mathcal{M}}{IJ}{\mu_1 \cdots \mu_{r_I}}{\nu_1 \cdots \nu_{s_I}}\tensor{}{_{\sigma_1 \cdots \sigma_{r_J}}^{\lambda_1 \cdots \lambda_{s_J}}} \, \multifield{\Psi}{J}{\sigma_1 \cdots \sigma_{r_J}}{\lambda_1 \cdots \lambda_{s_J}}\,.
\end{equation}

According to the notation we adopt for $\multifield{\Psi}{I}{A_I}{}$, $\multifield{\mathcal{M}}{IJ}{A_I}{B_J}$, $\multifield{\mathcal{F}}{IJ}{\alpha A_I}{B_J}$ and $\multifield{\mathcal{K}}{IJ}{\alpha\beta A_I}{B_J}$ are operators mapping tensors from $\Pi(r_J, s_J)$ to $\Pi(r_I, s_I)$, $\Pi(r_I, s_I) \otimes T_\mathscr{P} \mathscr{M}$ and $\Pi(r_I, s_I) \otimes T_\mathscr{P} \mathscr{M} \otimes T_\mathscr{P} \mathscr{M}$, respectively. As such, for $I = J$, the proposed system of equations can only be used if the rank of the equation of motion for a given field is the same as the rank of the field itself. In particular, the formalism we derive herein cannot be used to describe the Faraday tensor through the usual first-order Maxwell equations directly (cf. Subsec. \ref{subsec:no-coupling-fcg-kinetic-vanishing-friction}), which is addressed in references such as \cite{Santana2020}. Furthermore, when $A_I=B_J$, \textit{i.e.}, the $I$-th field has the same covariant and contravariant indices of the $J$-th field, the indices below each operator permit the distinction between otherwise ambiguous functions. For example, for a given pair $(I,J)$, $\multifield{\Psi}{I}{A_I}{} = \multifield{\Psi}{I}{\alpha}{}$ and $\multifield{\Psi}{J}{A_J}{} = \multifield{\Psi}{J}{\alpha}{}$, we can still have $\multifield{\mathcal{L}}{IJ}{A_I}{B_J} = \multifield{\mathcal{L}}{IJ}{\alpha}{\beta}\neq \multifield{\mathcal{L}}{JI}{\alpha}{\beta} = \multifield{\mathcal{L}}{JI}{A_I}{B_J}$. In most cases (even though this is not necessary), Eqs.~(\ref{eq:field-equation-operator}) may be understood as coming from the variation of the action of the given theory with respect to the $I$-th field. 

Let us then propose the following Ans\"atze:
\begin{subequations}
\begin{equation}
    \label{eq:initial-form-ansatze}
    \multifield{\Psi}{I}{A_I}{}(x, \epsilon_I) = \multifieldamp{\psi}{I}{A_I}{}(x, \epsilon_I) \, e^{iS_I(x)/\epsilon_I} \,,
\end{equation}
where
\begin{equation}
    \label{eq:initial-form-amplitude}
    \multifieldamp{\psi}{I}{A_I}{}(x, \epsilon_I) \leqdef \left[ \, \sum_{p = 0}^{N_I} \multifieldamp{\psi}{I}{A_I}{(p)}(x) \, \left(\frac{\epsilon_I}{i}\right)^p \, \right]\,, \quad N_I \ge 0 \,.
\end{equation}
\end{subequations}

\noindent Here, each tensor field is assumed to have its own set of complex-valued tensorial \textit{amplitude components} $\left\{\multifieldamp{\psi}{I}{A_I}{(0)}, ..., \multifieldamp{\psi}{I}{A_I}{(N_I)}\right\}$, grouped in the formal polynomial of Eq.~(\ref{eq:initial-form-amplitude}) to constitute the field \textit{amplitude}, a real-valued scalar \textit{phase} $S_I(x)$, and a real-valued positive \textit{smallness parameter} $\epsilon_I$, a dimensionless quantity proportional to the wavelength seen by a reference frame in the open set $\mathscr{O}$ where that field is defined. The gradient fields
\begin{equation}
	\label{eq:gradient-phase}
	\multifield{q}{I}{}{\alpha} \leqdef \partial_\alpha S_I
\end{equation}
are assumed to be nonzero everywhere on $\mathscr{O}$ \cite{Perlick2000}, such that the integral curves of each $\multifield{q}{I}{}{\alpha}$ form a congruence there in the open region of spacetime we are interested in, \textit{i.e.}, through each point passes one, and only one such curve. Naturally, despite not appearing in the above equations, we have to take the real part of the right-hand side (RHS) in each JWKB Ansatz to obtain the actual fields. Also, to distinguish between the full field and its amplitude components, we use the same kernel letter but uppercase for the former, and lowercase for the latter. By analogy with what we can consider for a scalar wave in Minkowski spacetime, we assume that there exist length scales
\begin{equation}
	\label{eq:amplitude-scale}
	\multifield{L}{I}{}{\psi} \stackrel{*}{\leqdef} \underset{\alpha, A_I, B_I}{\text{min}} \left(\frac{|\multifieldamp{\psi}{I}{A_I}{}(x)|}{|\partial_\alpha \multifieldamp{\psi}{I}{B_I}{}(x)|}\right)
\end{equation}
and
\begin{equation}
	\label{eq:q-wavevector-length-scale}
	\multifield{L}{I}{}{q} \stackrel{*}{\leqdef} \underset{\alpha, \beta, \gamma}{\text{min}} \left(\frac{|\multifield{q}{I}{}{\alpha}(x)|}{|\partial_\beta \multifield{q}{I}{}{\gamma}(x)|}\right)\,,
\end{equation} 
which, respectively, represent the typical lengths of variation for the amplitude and phase of every field $I$. Also, in a general curved spacetime, metric, torsion, and nonmetricity play the roles of ``refractive indices'' of sorts \cite{Kravtsov1990}, introducing three length scales:
\begin{equation}
	\label{eq:metric-length-scale}
	L_g \stackrel{*}{\leqdef} \underset{\alpha, \beta, \gamma, \lambda, \sigma}{\text{min}} \left(\frac{|g_{\alpha \beta}(x)|}{|\partial_\gamma g_{\lambda\sigma}(x)|} \right) \,,
\end{equation}
\begin{equation}
	\label{eq:torsion-length-scale}
	L_T \stackrel{*}{\leqdef} \underset{\alpha, \beta, \gamma, \lambda, \sigma, \mu, \nu}{\text{min}} \left(\frac{|\tensor{T}{^\alpha_{\beta\gamma}}(x)|}{|\partial_\lambda \tensor{T}{^\sigma_{\mu \nu}}(x)|} \right) \,,
\end{equation}
\begin{equation}
	\label{eq:nonmetricity-length-scale}
	L_Q \stackrel{*}{\leqdef} \underset{\alpha, \beta, \gamma, \lambda, \sigma, \mu, \nu}{\text{min}} \left(\frac{|\tensor{Q}{_{\alpha\beta\gamma}}(x)|}{|\partial_\lambda \tensor{Q}{_{\sigma\mu\nu}}(x)|} \right) \,,
\end{equation} 
which, together with $\multifield{L}{I}{}{\psi}$ and $\multifield{L}{I}{}{q}$, are to be taken into account when characterizing the desired regime of validity for the proposed solution (\ref{eq:rewritten-form-ansatze}). We then define the smallness parameter of field $I$ as $\epsilon_I \leqdef \lambda_I/L_I$, where $\lambda_I$ is its wavelength measured by a reference frame $u^\mu(x)$ \cite{Sachs1977}, and $L_I \leqdef \text{min}\{\multifield{L}{I}{}{\psi}, \multifield{L}{I}{}{q}, L_g, L_T, L_Q\}$, such that there are scenarios where $\epsilon_I\stackrel{*}{\ll} 1$, \textit{i.e.}, $\lambda_I \stackrel{*}{\ll} L_I$. It is worth mentioning that the characteristic length scales $\{\multifield{L}{I}{}{\psi}, \multifield{L}{I}{}{q}, L_g, L_T, L_Q\}$ are loosely defined by Eqs.~(\ref{eq:amplitude-scale}) through (\ref{eq:nonmetricity-length-scale}), respectively, but it is also common to see some of them appearing in the form of tensor quantities depending on the application under investigation. Indeed, for Lorentzian spacetimes, $L_g$ may be identified as one of the following curvature related scalars: $\widehat{R}^{-1/2}$, $(\widehat{R}_{\mu\nu} \widehat{R}^{\mu\nu})^{-1/4}$ or $(\widehat{R}_{\mu\nu\sigma\rho} \widehat{R}^{\mu\nu\sigma\rho})^{-1/4}$, where $\widehat{R}$, $\widehat{R}_{\mu\nu}$ and $\widehat{R}_{\mu\nu\sigma\rho}$ are, respectively, the Ricci scalar, the Ricci tensor, and the Riemann tensor. For spacetimes with $\widehat{R}_{\mu\nu} \ne 0$, the first two choices are reasonable candidates, whereas in the case, for example, of gravitational waves considered up to linear order on a Ricci-flat background, the latter is a better measurement of curvature, since the Kretschmann scalar $\widehat{R}_{\mu\nu\sigma\rho} \widehat{R}^{\mu\nu\sigma\rho}$ is the only nonvanishing quantity in that case \cite{Cusin2019, Lobato2021}.

While including different smallness parameters for distinct fields is surely a more general approach to follow, for simplicity, we choose to reexpress all Ans\"atze in terms of a single control parameter. We thus use Eqs.~(\ref{eq:initial-form-ansatze}) and (\ref{eq:initial-form-amplitude}) as an inspiration to consider alternative Ans\"{a}tze expressed in terms of a unique smallness parameter, $\epsilon$:
\begin{subequations}
\label{eqs:full-rewritten-ansatze}
\begin{equation}
	\label{eq:rewritten-form-ansatze}
	\multifield{\Psi}{I}{A_I}{}(x, \epsilon) = \multifieldamp{\psi}{I}{A_I}{}(x, \epsilon) \, e^{iS_I(x)/\epsilon} \,,
\end{equation}
with
\begin{equation}
	\label{eq:rewritten-form-amplitude}
	\multifieldamp{\psi}{I}{A_I}{}(x, \epsilon) \leqdef \left[ \, \sum_{p = 0}^{N} \multifieldamp{\psi}{I}{A_I}{(p)}(x) \, \left(\frac{\epsilon}{i}\right)^p \, \right]\,, \quad N \ge 0\,,
\end{equation}
where
\begin{equation}
	\label{eq:lowercase-n-uppercase-n}
	N \leqdef \underset{I}{\text{max}} \, \{N_I\} \,,
\end{equation}
\end{subequations}
and $\multifieldamp{\psi}{I}{A_I}{(p)} = 0$ for $p > N_I$ if $N_I < N$. In these alternative Ans\"atze, we have ordered the fields $\multifield{\Psi}{I}{A_I}{}$ such that the $M$-tuple $(\epsilon_1, \epsilon_2, ..., \epsilon_M)$ satisfies $\epsilon_1 \le \epsilon_2 \le \cdots \le \epsilon_M$, and $\epsilon$ is supposed to lie somewhere in the interval $[\epsilon_1, \epsilon_M]$. As such, if $\epsilon$ is small, all fields with $\epsilon_I \le \epsilon$ will require only equations involving their leading-order amplitudes $\multifieldamp{\psi}{I}{A_I}{(0)}$ to have most of their dynamics described, a situation which we refer to as the \textit{geometrical optics regime} (GO), whereas the description of other fields may demand that higher-order amplitude components $\multifieldamp{\psi}{I}{A_I}{(p)}$ ($p = 1, ..., P_I > 0$) should be taken into account, a broader situation which we refer to as the \textit{diffraction regime} or \textit{beyond geometrical optics} (BGO). With these transformations, the \textit{wave vector} associated with field $I$ is given by
\begin{equation}
	\label{eq:actual-wave-vector}
	\multifield{k}{I}{}{\mu} \leqdef \epsilon^{-1} \, \multifield{q}{I}{}{\mu} = \epsilon^{-1} \, \partial_\mu S_I \,.
\end{equation}

Substituting Eq.~(\ref{eq:rewritten-form-ansatze}) into Eq.~(\ref{eq:field-equation}) and demanding its validity for each order of $\epsilon$, we derive the following set of equations:
\begin{subequations}
\label{eqs:hierarchy-general}
\begin{itemize}
\item \textit{dominant $\epsilon^{-2}$ order}:
\begin{equation}
    \label{eq:leading-order-general}
    \sum_{J=1}^M \, \multifield{\mathcal{D}}{IJ}{A_I}{B_J} \, \multifieldamp{\psi}{J}{B_J}{(0)} \, e^{iS_J/\epsilon} = 0 \,,
\end{equation}
\item \textit{first subdominant $\epsilon^{-1}$ order}:
\begin{equation}
    \label{eq:1st-subleading-order-general}
    \sum_{J=1}^M \, \left[\, \multifield{\mathcal{D}}{IJ}{A_I}{B_J} \, \multifieldamp{\psi}{J}{B_J}{(1)} + \multifield{\mathcal{T}}{IJ}{A_I}{B_J} \, \multifieldamp{\psi}{J}{B_J}{(0)} \,\right] \, e^{iS_J/\epsilon} = 0 \,,
\end{equation}
\item \textit{remaining subdominant $\epsilon^{p}$ order}:
\begin{equation}
    \label{eq:higher-subleading-order-general}
    \sum_{J=1}^M \, \left[\, \multifield{\mathcal{D}}{IJ}{A_I}{B_J} \, \multifieldamp{\psi}{J}{B_J}{(p+2)} + \multifield{\mathcal{T}}{IJ}{A_I}{B_J} \, \multifieldamp{\psi}{J}{B_J}{(p+1)} + \multifield{\mathcal{L}}{IJ}{A_I}{B_J} \, \multifieldamp{\psi}{J}{B_J}{(p)} \,\right] \, e^{iS_J/\epsilon} = 0 \,, \quad (0 \le p \le N)\,.
\end{equation}
\end{itemize}
\end{subequations}

Here, we have introduced the operators
\begin{equation}
	\label{eq:dispersion-operator}
	\multifield{\mathcal{D}}{IJ}{A_I}{B_J} \leqdef \multifield{\mathcal{K}}{IJ}{\alpha\beta A_I}{B_J} \, \multifield{q}{J}{}{\alpha} \, \multifield{q}{J}{}{\beta} \,,
\end{equation}
\begin{equation}
	\label{eq:transport-operator}
	\multifield{\mathcal{T}}{IJ}{A_I}{B_J} \leqdef \multifield{\mathcal{K}}{IJ}{\alpha\beta A_I}{B_J} \, \multifield{D}{J}{}{\alpha\beta} + \multifield{\mathcal{F}}{IJ}{\alpha A_I}{B_J} \, \multifield{q}{J}{}{\alpha} \,,
\end{equation}
where
\begin{equation}
	\label{eq:wave-vector-and-wave-vector-transport-operator}
	\multifield{D}{I}{}{\alpha\beta} \leqdef \multifield{q}{I}{}{\alpha} \, \nabla_\beta + \multifield{q}{I}{}{\beta} \, \nabla_\alpha + \nabla_\alpha \multifield{q}{I}{}{\beta} \,.
\end{equation}

$\multifield{D}{I}{}{\alpha\beta}$ is a derivative operator associated to $\multifield{q}{I}{}{\alpha}$ which will play an important role in the transport equations derived below for the various $\multifieldamp{\psi}{I}{A_I}{(p)}$ appearing in Eq.~(\ref{eq:rewritten-form-ansatze}). Eq.~(\ref{eq:dispersion-operator}) defines what we refer to as \textit{dispersion tensors} $\multifield{\mathcal{D}}{IJ}{A_I}{B_J}$ for $\multifield{q}{I}{}{\alpha}$. In particular, the following simplifications follow if $\multifield{\mathcal{D}}{IJ}{A_I}{B_J} = 0$:

\begin{subequations}
\begin{itemize}
\item \textit{dominant $\epsilon^{-2}$ order}:
\begin{equation}
    \label{eq:vanishing-dispersion-operators}
    \multifield{\mathcal{K}}{IJ}{\alpha\beta A_I}{B_J} \, \multifield{q}{J}{}{\alpha} \, \multifield{q}{J}{}{\beta} = 0\,.
\end{equation}
\item \textit{first subdominant $\epsilon^{-1}$ order}:
\begin{equation}
    \label{eq:1st-subleading-order-simplified}
    \sum_{J=1}^M \, \multifield{\mathcal{T}}{IJ}{A_I}{B_J} \, \multifieldamp{\psi}{J}{B_J}{(0)} \, e^{iS_J/\epsilon} = 0 \,,
\end{equation}
\item \textit{remaining subdominant $\epsilon^{p}$ order}:
\begin{equation}
    \label{eq:higher-subleading-order-simplified}
    \sum_{J=1}^M \, \left[\, \multifield{\mathcal{T}}{IJ}{A_I}{B_J} \, \multifieldamp{\psi}{J}{B_J}{(p+1)} + \multifield{\mathcal{L}}{IJ}{A_I}{B_J} \, \multifieldamp{\psi}{J}{B_J}{(p)} \,\right] \, e^{iS_J/\epsilon} = 0 \,, \quad (p \ge 0) \,.
\end{equation}
\end{itemize}
\end{subequations}

If this is the case, from the functional dependence of the tensors $\multifield{\mathcal{D}}{IJ}{A_I}{B_J}$, $\multifield{\mathcal{T}}{IJ}{A_I}{B_J}$ and $\multifield{\mathcal{L}}{IJ}{A_I}{B_J}$ on $\multifield{q}{I}{}{\alpha}$ and on covariant derivatives, Eqs.~(\ref{eq:1st-subleading-order-simplified}) and (\ref{eq:higher-subleading-order-simplified}) can be understood as a system of \textit{perturbative} evolution equations for the different order amplitudes of all fields involved, in which the behavior of $\multifieldamp{\psi}{I}{A_I}{(p+1)}$ is influenced only by $\multifieldamp{\psi}{J}{A_I}{(p)}$, not by $\multifieldamp{\psi}{J}{A_I}{(p+2)}$. Furthermore, the results concerning the leading-order amplitudes $\multifieldamp{\psi}{J}{A_J}{(0)}$ are indifferent to amplitudes beyond geometrical optics being present or not. Finally, in the special case in which all fields share a single phase, \textit{i.e.}, $S_I(x) = S(x) \, \, (1 \le I \le M)$, but not necessarily $\multifield{\mathcal{D}}{IJ}{A_I}{B_J} = 0$ holds, the exponentials appearing in Eqs.~(\ref{eq:leading-order-general}), (\ref{eq:1st-subleading-order-general}) and (\ref{eq:higher-subleading-order-general}) may be naturally factored out. 

\section{Decomposition of the kinetic tensor}
\label{sec:decomposition-kinetic-operator}

While the above results are quite general and independent of the tensor field(s) considered, it is interesting to start extracting information from Eqs.~(\ref{eq:leading-order-general}) through (\ref{eq:higher-subleading-order-general}) by performing the following decomposition of the kinetic tensor $\multifield{\mathcal{K}}{IJ}{\alpha\beta A_I}{B_J}$:
\begin{subequations}
\begin{equation}
   \label{eq:kinetic-operator-decomposition}
   \multifield{\mathcal{K}}{IJ}{\alpha\beta A_I}{B_J} \reqdef \delta_{IJ} \, \left(\multifield{\overset{(1)}{\kappa}}{I}{A_I}{B_I} \, g^{\alpha\beta} + \multifield{\overset{(2)}{\kappa}}{I}{\alpha\beta A_I}{B_I} \right) + \multifield{\overset{(3)}{\kappa}}{IJ}{\alpha\beta A_I}{B_J}\,,
\end{equation}
satisfying the constraints
\begin{equation}
    \label{eq:kinetic-operator-decomposition-constraints}
    \multifield{\overset{(1)}{\kappa}}{I}{A_I}{B_I} \leqdef \frac{1}{4} g_{\alpha\beta} \, \multifield{\mathcal{K}}{II}{\alpha\beta A_I}{B_I} \,, \quad g_{\alpha\beta} \, \multifield{\overset{(2)}{\kappa}}{I}{\alpha\beta A_I}{B_I} \leqdef 0 \,, \quad \multifield{\overset{(3)}{\kappa}}{II}{\alpha\beta A_I}{B_I} \leqdef 0\,.    
\end{equation}
\end{subequations}

\noindent This splits the kinetic tensor into a part that does not mix different fields (\textit{field-diagonal} or F-diagonal), represented by the term within round brackets, and a remainder which gives the coupling between different fields (the kinetic terms in the $I$-th equation of motion that depend on the $J$-th field, with $J \ne I$), and allows us to identify in what follows the conditions leading to the well-known results of GO and BGO in literature \cite{Harte2013, Cusin2019}. Furthermore, the F-diagonal term is additionally split in two, one proportional to $g^{\alpha\beta}$ (\textit{metric-diagonal} or G-diagonal), which generally leads to the d'Alembertian operator appearing in wave-like equations of motion for various theories (cf. Sec. \ref{sec:examples-lorentzian-spacetimes}), and a remainder, which is only F-diagonal. We now use this decomposition to derive the results of having simpler types of kinetic tensors isolatedly.

\subsection{FG-diagonal kinetic tensor}
\label{subsec:fg-diagonal-kinetic-operator} 

First, we consider an \textit{FG-diagonal} kinetic tensor, \textit{i.e.}, $\multifield{\overset{(1)}{\kappa}}{I}{A_I}{B_I} \ne 0$, but $\multifield{\overset{(2)}{\kappa}}{I}{\alpha\beta A_I}{B_I} = \multifield{\overset{(3)}{\kappa}}{IJ}{\alpha\beta A_I}{B_J} = 0$. Then, Eq.~(\ref{eq:leading-order-general}) gives 
\begin{equation}
	\label{eq:dispersion-fg-diagonal}
	\mathcal{C}_{I} \, \multifield{\overset{(1)}{\kappa}}{I}{A_I}{B_I} \, \multifieldamp{\psi}{I}{B_I}{(0)} = 0\,, \quad \mathcal{C}_{I} \leqdef g^{\alpha\beta} \, \multifield{q}{I}{}{\alpha} \, \multifield{q}{I}{}{\beta}\,,
\end{equation}
where we factored out the phase for the $I$-th field due to the F-diagonality, and introduced $\mathcal{C}_{I}$, which attests to the \textit{character} (timelike, lightlike, or spacelike) of the wave vectors $\multifield{q}{I}{}{\alpha}$. If $\multifieldamp{\psi}{I}{A_I}{(0)} 
 \notin \text{ker}\left\{\multifield{\overset{(1)}{\kappa}}{I}{A_I}{B_I}\right\}$, Eq.~(\ref{eq:dispersion-fg-diagonal}) leads to
\begin{equation}
    \label{eq:null-condition}
    \mathcal{C}_{I} = g^{\alpha\beta} \, \multifield{q}{I}{}{\alpha} \, \multifield{q}{I}{}{\beta} = 0 \Rightarrow \multifield{\mathcal{D}}{IJ}{A_I}{B_J} = 0\,.
\end{equation}

The gradient of the above constraint gives \cite{Santana2017}
\begin{equation}
    \label{eq:wave-vector-transport}
	\multifield{q}{I}{\alpha}{} \, \nabla_\alpha \multifield{q}{I}{\mu}{} = \left[\tensor{T}{_{\alpha\beta}^\mu} + (1/2) \, \tensor{Q}{_{\alpha\beta}^\mu} - \tensor{Q}{^\mu_{\alpha\beta}} \right] \, \multifield{q}{I}{\alpha}{} \multifield{q}{I}{\beta}{} \,.
\end{equation}

Therefore, an FG-diagonal kinetic tensor implies that the integral curves of $\multifield{q}{I}{\alpha}{}$ are null and satisfy the transport equations given by (\ref{eq:wave-vector-transport}). For Lorentzian theories of gravity or metric-affine theories with a metric-compatible connection and a completely antisymmetric torsion, this transport equation implies that the integral curves of $\multifield{q}{I}{\alpha}{}$ are affinely parametrized (metric) geodesics and affinely parametrized autoparallels \cite{Santana2017}. For metric-affine spacetimes with a symmetric Weyl connection, Eq.~(\ref{eq:wave-vector-transport}) implies that the integral curves of $\multifield{q}{I}{\alpha}{}$ are affinely parametrized (metric) geodesics and nonaffinely parametrized autoparallels \cite{Santana2017}. Furthermore, the vanishing dispersion relations lead to Eqs.~(\ref{eq:1st-subleading-order-simplified}) and (\ref{eq:higher-subleading-order-simplified}), which read: 
\begin{subequations}
\begin{equation}
    \label{eq:fg-1st-subleading-order}
    \multifield{\overset{(1)}{\kappa}}{I}{A_I}{B_I} \, \multifield{\mathscr{D}}{I}{(g)}{} \, \multifieldamp{\psi}{I}{B_I}{(0)} + \sum_{J=1}^M \, \multifield{\mathcal{F}}{IJ}{\alpha A_I}{B_J} \, \multifield{q}{J}{}{\alpha} \, \multifieldamp{\psi}{J}{B_J}{(0)} \, e^{i(S_J - S_I)/\epsilon} = 0 \,,
\end{equation}
and
\begin{equation}
    \label{eq:fg-higher-subleading-order}
    \multifield{\overset{(1)}{\kappa}}{I}{A_I}{B_I} \, \multifield{\mathscr{D}}{I}{(g)}{} \, \multifieldamp{\psi}{I}{B_I}{(p + 1)} + \sum_{J=1}^M \, \multifield{\mathcal{F}}{IJ}{\alpha A_I}{B_J} \, \multifield{q}{J}{}{\alpha} \, \multifieldamp{\psi}{J}{B_J}{(p + 1)} \, e^{i(S_J - S_I)/\epsilon} = -\sum_{J=1}^M \, \multifield{\mathcal{L}}{IJ}{A_I}{B_J} \, \multifieldamp{\psi}{J}{B_J}{(p)} \, e^{i(S_J - S_I)/\epsilon} \,,
\end{equation}    
\end{subequations}
for $0 \le p \le N$, where
\begin{equation}
	\label{eq:g-scalar-transport-operator}
	\multifield{\mathscr{D}}{I}{(g)}{} \leqdef g^{\alpha\beta} \, \multifield{D}{I}{}{\alpha\beta} = 2 \, \multifield{q}{I}{\alpha}{} \, \nabla_\alpha + \nabla_\alpha \multifield{q}{I}{\alpha}{} + \tensor{Q}{_{\beta\alpha}^\beta} \, \multifield{q}{I}{\alpha}{}\,.
\end{equation}

Eqs.~(\ref{eq:fg-1st-subleading-order}) and (\ref{eq:fg-higher-subleading-order}) are evolution equations for the different amplitudes $\multifieldamp{\psi}{I}{A_I}{(p)}$ of $\multifield{\Psi}{I}{A_I}{}$ along the integral curves of $\multifield{q}{I}{\alpha}{}$. Eqs.~(\ref{eq:null-condition}), (\ref{eq:wave-vector-transport}) and (\ref{eq:fg-1st-subleading-order}) provide the results leading to the well-known \textit{laws of geometrical optics} appearing in the literature \cite{Ehlers1967, Ehlers2022, Misner1973, Schneider1992, Perlick2000, Ellis2012} whereas the remaining orders allow us to go BGO. The GO results will be properly illustrated later in Secs. \ref{sec:examples-lorentzian-spacetimes} and \ref{sec:examples-metric-affine-spacetimes}. It is worth stressing that, for some choice of fields, the laws of geometrical optics are contingent on choosing a gauge condition, such as the Lorenz gauge for the electromagnetic potential $A^\alpha$ or the harmonic gauge for gravitational wave metric perturbations [cf. Subsec. (\ref{subsec:no-coupling-fcg-kinetic-vanishing-friction})].

\subsection{FC-diagonal kinetic tensor}
\label{subsec:fc-diagonal-kinetic-operators}

Now, of particular relevance is the case where the F-diagonal part of the kinetic tensor can be written as
\begin{equation}
	\label{eq:fc-diagonal-kinetic-operator}
	\multifield{\overset{(1)}{\kappa}}{I}{A_I}{B_I} \reqdef \multifield{\overset{(1)}{\varkappa}}{I}{}{} \, \delta^{A_I}_{B_I} \,, \quad \multifield{\overset{(2)}{\kappa}}{I}{\alpha\beta A_I}{B_I} \reqdef \multifield{\overset{(2)}{\varkappa}}{I}{\alpha\beta}{}{} \, \delta^{A_I}_{B_I}\,,
\end{equation}
where
\begin{equation}
    \delta^{A_I}_{B_I} \leqdef \delta^{\mu_1}_{\sigma_1} \cdots \delta^{\mu_{r_I}}_{\sigma_{r_I}} \delta_{\nu_1}^{\lambda_1} \cdots \delta_{\nu_{s_I}}^{\lambda_{s_I}}
\end{equation}
is the identity operator in the tensor space $\Pi_{\mathscr{P}}(r_I,s_I)$ for every $\mathscr{P} \in \mathscr{M}$. We refer to this possibility as \textit{FC-diagonality}, since the F-diagonal part of the kinetic tensor does not mix different components of a given field. Then, Eq.~(\ref{eq:leading-order-general}) gives
\begin{equation}
	\label{eq:fc-dispersion-diagonal}
	\left(\multifield{\overset{(1)}{\varkappa}}{I}{}{} \, g^{\alpha\beta} + \multifield{\overset{(2)}{\varkappa}}{I}{\alpha\beta}{} \right) \, \multifield{q}{I}{}{\alpha} \, \multifield{q}{I}{}{\beta} \, \multifieldamp{\psi}{I}{A_I}{(0)} = 0\,.
\end{equation}

In this case, if $\multifieldamp{\psi}{I}{A_I}{(0)}$ vanishes at most in isolated points,
\begin{equation}
	\label{eq:deviation-nullity-condition}
	\left(\multifield{\overset{(1)}{\varkappa}}{I}{}{} \, g^{\alpha\beta} + \multifield{\overset{(2)}{\varkappa}}{I}{\alpha\beta}{} \right) \, \multifield{q}{I}{}{\alpha} \, \multifield{q}{I}{}{\beta} = 0 \Rightarrow \multifield{\mathcal{D}}{IJ}{A_I}{B_J} = 0 \,,
\end{equation}
a nontrivial example of vanishing dispersion tensors [cf. Eq.~(\ref{eq:vanishing-dispersion-operators})]. In addition,  Eq.~(\ref{eq:transport-operator}) can be rewritten as
\begin{equation}
	\label{eq:fc-transport-operator}
	\multifield{\mathcal{T}}{IJ}{A_I}{B_J} \overset{*}{=} \delta_{IJ} \, \delta^{A_I}_{B_I} \, \left( \multifield{\overset{(1)}{\varkappa}}{I}{}{} \, \multifield{\mathscr{D}}{I}{(g)}{} + \multifield{\mathscr{D}}{I}{(\varkappa)}{} \right) + \multifield{\mathcal{F}}{IJ}{\alpha A_I}{B_J} \, \multifield{q}{J}{}{\alpha} \,, \quad  \multifield{\mathscr{D}}{I}{(\varkappa)}{} \leqdef \multifield{\overset{(2)}{\varkappa}}{I}{\alpha\beta}{} \, \multifield{D}{I}{}{\alpha\beta} \,,
\end{equation}
so that Eqs.~(\ref{eq:1st-subleading-order-simplified}) and (\ref{eq:higher-subleading-order-simplified}) give
\begin{subequations}
\begin{equation}
    \label{eq:fc-1st-subleading-order}
    \left(\multifield{\mathscr{D}}{I}{(g)}{} + \frac{\multifield{\mathscr{D}}{I}{(\varkappa)}{}}{\multifield{\overset{(1)}{\varkappa}}{I}{}{}} \right) \multifieldamp{\psi}{I}{A_I}{(0)} + \sum_{J=1}^M \, \frac{\multifield{\mathcal{F}}{IJ}{\alpha A_I}{B_J} \, \multifield{q}{J}{}{\alpha}}{\multifield{\overset{(1)}{\varkappa}}{I}{}{}} \, \multifieldamp{\psi}{J}{B_J}{(0)} \, e^{i(S_J - S_I)/\epsilon} = 0 \,,
\end{equation}
and
\begin{equation}
    \label{eq:fc-higher-subleading-order}
    \left(\multifield{\mathscr{D}}{I}{(g)}{} + \frac{\multifield{\mathscr{D}}{I}{(\varkappa)}{}}{\multifield{\overset{(1)}{\varkappa}}{I}{}{}} \right) \multifieldamp{\psi}{I}{A_I}{(p+1)} + \sum_{J=1}^M \, \frac{\multifield{\mathcal{F}}{IJ}{\alpha A_I}{B_J} \, \multifield{q}{J}{}{\alpha}}{\multifield{\overset{(1)}{\varkappa}}{I}{}{}} \, \multifieldamp{\psi}{J}{B_J}{(p+1)} \, e^{i(S_J - S_I)/\epsilon} = -\sum_{J=1}^M \, \frac{\multifield{\mathcal{L}}{IJ}{A_I}{B_J}}{\multifield{\overset{(1)}{\varkappa}}{I}{}{}} \, \multifieldamp{\psi}{J}{B_J}{(p)} \, e^{i(S_J - S_I)/\epsilon} \,,
\end{equation}
\end{subequations}
for $0 \le p \le N$.

\subsection{FCG-diagonal kinetic tensor}
\label{subsec:fcg-diagonal-kinetic-operators}

If $\multifield{\overset{(2)}{\varkappa}}{I}{\alpha\beta}{} = 0$, the kinetic tensor is FCG-diagonal, implying that $\mathcal{C}_{I} = 0$ and that $\multifield{q}{I}{\alpha}{}$ satisfies Eq.~(\ref{eq:wave-vector-transport}). In turn, Eqs.~(\ref{eq:1st-subleading-order-simplified}) and (\ref{eq:higher-subleading-order-simplified}) simplify to the following:
\begin{subequations}
\begin{equation}
    \label{eq:fcg-1st-subleading-order}
	\multifield{\mathscr{D}}{I}{(g)}{} \multifieldamp{\psi}{I}{A_I}{(0)} + \sum_{J=1}^M \, \frac{\multifield{\mathcal{F}}{IJ}{\alpha A_I}{B_J} \, \multifield{q}{J}{}{\alpha}}{\multifield{\overset{(1)}{\varkappa}}{I}{}{}} \, \multifieldamp{\psi}{J}{B_J}{(0)} \, e^{i(S_J - S_I)/\epsilon} = 0 \,,
\end{equation}
and
\begin{equation}
    \label{eq:fcg-higher-subleading-order}
	\multifield{\mathscr{D}}{I}{(g)}{} \multifieldamp{\psi}{I}{A_I}{(p+1)} + \sum_{J=1}^M \, \frac{\multifield{\mathcal{F}}{IJ}{\alpha A_I}{B_J} \, \multifield{q}{J}{}{\alpha}}{\multifield{\overset{(1)}{\varkappa}}{I}{}{}} \, \multifieldamp{\psi}{J}{B_J}{(p+1)} \, e^{i(S_J - S_I)/\epsilon} = -\sum_{J=1}^M \, \frac{\multifield{\mathcal{L}}{IJ}{A_I}{B_J}}{\multifield{\overset{(1)}{\varkappa}}{I}{}{}} \, \multifieldamp{\psi}{J}{B_J}{(p)} \, e^{i(S_J - S_I)/\epsilon} \,,
\end{equation}
\end{subequations}
for $0 \ge p \ge N$. As we explore in the following examples, these simplified transport equations are common when describing the evolution of different fields.
\begin{figure}
    \centering
    \includegraphics[scale = 0.7]{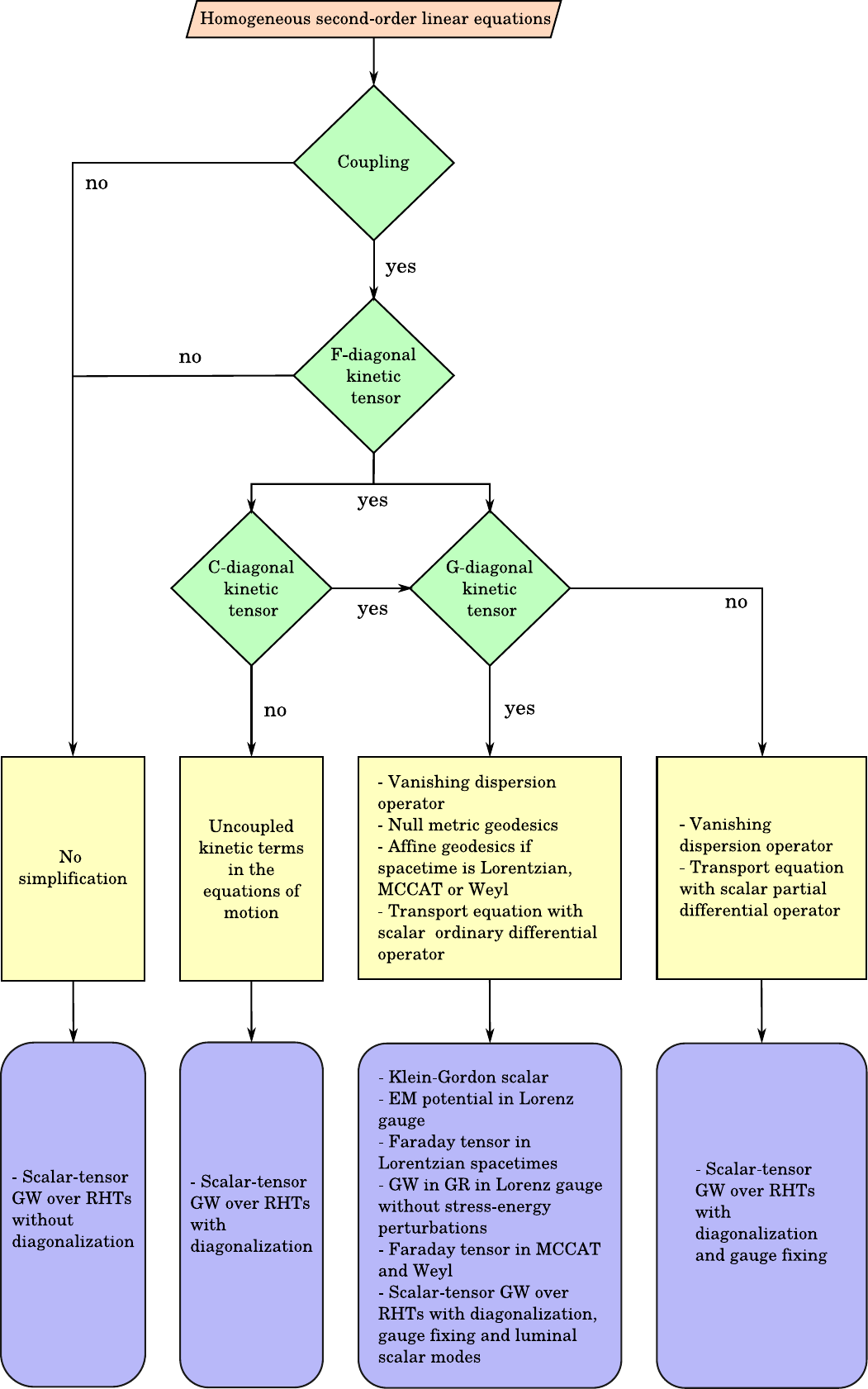}
    \caption{Flow diagram depicting the sequence of conditions (green diamonds) satisfied by the original universe of equations of motion (orange parallelograms) leading to relevant results (yellow rectangles) and corresponding examples (blue parallelograms with round vertices). The latter are presented in Secs.~\ref{sec:examples-lorentzian-spacetimes} and \ref{sec:examples-metric-affine-spacetimes}. This diagram does not exhaust all possibilities of conditional tests, since F-diagonality, C-diagonality and G-diagonality could, in principle, be tested independently, rather than displayed in serialized form. As such, this flowchart is just a visual representation of interesting combinations among the three diagonality conditions that lead to the relevant universal results presented in Sec.~\ref{sec:decomposition-kinetic-operator}.}
    \label{fig:flow-chart}
\end{figure}

\section{Examples in Lorentzian spacetimes}
\label{sec:examples-lorentzian-spacetimes}

To illustrate the extent of the results we have just derived, let us now consider several examples of fields whose dynamics is described by a wave-like equation in the form of Eq.~(\ref{eq:field-equation}). We start by considering several fields on a prescribed Lorentzian spacetime which evolve in vacuum (only under the influence of gravity). Also, by noting that the leading-order and first subleading-order results ($p = n, n+1$) have equal powers of the wave vectors $\multifield{q}{I}{}{\alpha}$ appearing in both sides of every equation, we can substitute $\multifield{q}{I}{}{\alpha}$ by $\multifield{k}{I}{}{\alpha}$, and recast our derivations in terms of the actual physical wave vector $\multifield{k}{I}{}{\alpha}$.

\subsection{No coupling, FCG-diagonal kinetic tensors and vanishing friction tensors}
\label{subsec:no-coupling-fcg-kinetic-vanishing-friction}

First, we consider noncoupled fields in Lorentzian spacetimes, with examples being the Klein-Gordon scalar $\Phi$, the electromagnetic potential vector $A_\mu$, the Faraday tensor $F_{\mu\nu}$, and a trace-reversed gravitational wave (GW) perturbation in general relativity $C_{\mu\nu}$. A thorough study of these fields can be found in various classical texts \cite{Misner1973, Schneider1992, Ellis2012} as well as in more recent works \cite{Harte2019, Santana2020}. Their second-order equations of motion read, respectively:
\begin{equation}
	\label{eq:klein-gordon-equation}
	\widehat{\square} \, \Phi - \left(\xi \widehat{R} + \mu^2\right) \Phi = 0\,,
\end{equation}
\begin{equation}
	\label{eq:electromagnetic-potential-equation}
	\widehat{\square} \, A_\mu - \tensor{\widehat{R}}{_\mu^\nu} \, A_\nu = 0\,, \quad \left(\hatnabla_\alpha A^\alpha = 0\right)\,,
\end{equation}
\begin{equation}
	\label{eq:faraday-tensor-equation}
	\widehat{\square} \, F_{\mu\nu} + \tensor{\widehat{R}}{_{\mu\nu}^{\lambda\sigma}} \, F_{\lambda\sigma} + 2 \tensor{\widehat{R}}{_{[\mu}^\lambda} \, F_{\nu]\lambda} = 0\,, \quad \left(\hatnabla_\nu F^{\mu\nu} = 0\,, \hatnabla_{[\lambda} F_{\mu\nu]} = 0\right)\,,
\end{equation}
and
\begin{equation}
\label{eq:gr-gw-field-equation}
   \nunderbar{\widehat{\square}} C_{\mu\nu} + \left(2 \nunderbar{\widehat{R}}\tensor{}{^\lambda_\mu^\sigma_\nu} - 2 \nunderbar{\widehat{R}}\tensor{}{^\lambda_{(\mu}} \, \delta_{\nu)}^\sigma - \munderbar{g}_{\mu\nu} \, \nunderbar{\widehat{R}}\tensor{}{^{\lambda\sigma}} + \nunderbar{\widehat{R}} \, \delta_\mu^\lambda \delta_\nu^\sigma \right) C_{\lambda\sigma} = 0 \,, \quad \left(\munderbar{\hatnabla}^\nu C_{\mu\nu} = 0\right) \,.
\end{equation}

We have included auxiliary gauge conditions inside parentheses to remind ourselves that they must be further enforced to constrain the general solutions of the second-order equations of motion. Particularly, in the case of $F_{\mu\nu}$, which we describe by its second-order wave equation in Lorentzian spacetimes, first-order Maxwell equations act as gauge conditions of sorts, having to be later imposed in order to remove nonphysical solutions. In the equations above, $\,\widehat{\cdot}\,$ represents $\,\cdot\,$ computed using the Levi-Civita connection (since we are in the context of Lorentzian spacetimes), whereas $\,\nunderbar{\cdot}\,$ denotes $\,\cdot\,$ computed on a background gravitational field, which in this case acts as the prescribed metric [cf. the comment after Eq.~(\ref{eq:field-equation-operator})].

A quick inspection shows that all of these fields obey equations of motion possessing FCG-diagonal kinetic tensors, vanishing friction tensors, and mass tensors differing according to the field rank. Table \ref{tab:examples-lorentzian-spacetimes} summarizes them in a compact form:
\begin{table}[h!]
    \renewcommand{\arraystretch}{2}
    \begin{center}
        \begin{tabular}{l|c|c|c}
            \hline
            \hline
            $\multifield{\Psi}{}{A}{}$ & $\multifield{\mathcal{K}}{}{\alpha\beta A}{B}$ & $\multifield{\mathcal{F}}{}{\alpha A}{B}$ & $\multifield{\mathcal{M}}{}{A}{B}$ \\
            \hline
            $\Phi$       & $g^{\alpha\beta}$ & $ 0 $ & $-\left(\xi \widehat{R} + \mu^2\right)$ 		\\
            $A_\mu$      & $\delta_\mu^\nu \, g^{\alpha\beta}$ & $0$  & $-\tensor{\widehat{R}}{_\mu^\nu}$ \\
            $F_{\mu\nu}$ & $\delta_{\mu}^\lambda \, \delta_{\nu}^\sigma \, g^{\alpha\beta}$ & $0$ & $\tensor{\widehat{R}}{_{\mu\nu}^{\lambda\sigma}} + 2 \tensor{\widehat{R}}{_{[\mu}^\lambda} \, \delta_{\nu]}^\sigma$ \\
            $C_{\mu\nu}$ & $\delta_{\mu}^\lambda \, \delta_{\nu}^\sigma \, g^{\alpha\beta}$ & $0$ & $2 \nunderbar{\widehat{R}}\tensor{}{^\lambda_\mu^\sigma_\nu} - \left(2 \nunderbar{\widehat{R}}\tensor{}{^\lambda_{(\mu}} \, \delta_{\nu)}^\sigma + \munderbar{g}_{\mu\nu} \, \nunderbar{\widehat{R}}\tensor{}{^{\lambda\sigma}}\right) + \nunderbar{\widehat{R}} \, \delta_\mu^\lambda \delta_\nu^\sigma$ \\
            \hline
            \hline
        \end{tabular}
    \end{center}
    \caption{Examples of fields which satisfy equations of motion with FCG-diagonal kinetic tensors, vanishing friction tensors and nonvanishing mass tensors. The $I$ index for different fields has been dropped due to the absence of coupling.}
    \label{tab:examples-lorentzian-spacetimes}
\end{table}

From this observation, by proposing JWKB Ans\"atze:
\begin{equation}
    \label{eq:general-ansatz-no-coupling}
    \multifield{\Psi}{}{A}{}(x, \epsilon) = \left[\,\sum_{p = 0}^N \multifieldamp{\psi}{}{A}{(p)}(x) \, \left(\frac{\epsilon}{i}\right)^p\,\right] \, e^{iS(x)/\epsilon}\,,
\end{equation}
and following the general procedure we have developed in the prior sections (notice that we have dropped the $I$ index due to the absence of coupling), the wave vectors $k_\mu$ associated to each Ansatz satisfy [cf. Eq.~(\ref{eq:null-condition})]:
\begin{equation}
    \label{eq:fcg-kinetic-vanishing-friction-null-geodesics-autoparallels}
    k_\mu \, k^\mu = 0 \Rightarrow k^\nu \hatnabla_\nu k^\mu = 0\,,
\end{equation}
\textit{i.e.}, their associated integral curves are \textit{null} and both \textit{affinely parametrized metric geodesics and autoparallels} [since the RHS of Eq.~(\ref{eq:wave-vector-transport}) vanishes for Lorentzian spacetimes]. Moreover, provided friction tensors vanish altogether, their leading-order amplitudes satisfy a simplified version of Eq.~(\ref{eq:fcg-1st-subleading-order}):
\begin{equation}
	\label{eq:fcg-kinetic-vanishing-friction-leading-order-transport}
	\multifield{\widehat{\mathscr{D}}}{}{(g)}{} \multifieldamp{\psi}{}{A}{(0)} = \left(2 \, \multifield{k}{}{\alpha}{} \hatnabla_\alpha + \hatnabla_\alpha \multifield{k}{}{\alpha}{}\right) \multifieldamp{\psi}{}{A}{(0)} = 0\,,
\end{equation}
where $\multifieldamp{\psi}{}{A}{(0)}$ is equal to $\multifieldamp{\phi}{}{}{(0)}$, $\multifieldamp{a}{}{}{(0)\mu}$, $\multifieldamp{f}{}{}{(0)\mu\nu}$, and $\multifieldamp{c}{}{}{(0)\mu\nu}$, respectively, \textit{i.e.}, the leading-order amplitudes of the JWKB Ans\"atze for the Klein-Gordon scalar, the electromagnetic vector potential, the Faraday tensor and the GW perturbation, respectively. Next, with the exception of the Klein-Gordon field (which is a scalar), the gauge conditions appearing inside parentheses for $A_\mu$, $F_{\mu\nu}$ and $C_{\mu\nu}$ imply that $\multifieldamp{\psi}{}{A}{(0)}$ are transverse to their corresponding wave vectors $k^\mu$. Therefore, since GO results do not depend on mass tensors, by knowing that all of the above fields obey equations of motion with FCG-diagonal kinetic tensors and vanishing friction tensors, they should all obey the following \textit{laws of geometrical optics}:
\begin{enumerate}[(i)]
    \item the associated wave vector $k^\mu$ is null and (affine and metric) geodesic, 
    \item the leading-order amplitude $\multifieldamp{\psi}{}{A}{(0)}$ is transverse to $k^\mu$ (with the exception of the Klein-Gordon scalar), and
    \item $\multifieldamp{\psi}{}{A}{(0)}$ evolves along the integral curves of $k^\mu$ according to Eq.~(\ref{eq:fcg-kinetic-vanishing-friction-leading-order-transport}). 
\end{enumerate}

Since Eq.~\ref{eq:fcg-kinetic-vanishing-friction-leading-order-transport} is a tensor equation, when we are describing light propagation in the geometrical optics limit, either using $A_\mu$ or $F_{\mu\nu}$ as the fundamental field, law (iii) can be recast as two indepedent transport laws for the intensity and the polarization of light. The former equation gives the pleasant result that a notion of ``photon number'' is preserved along the selected null geodesic \cite{Schneider1992, Ellis2012, Santana2020}. In turn, the latter transport law simplifies to a parallel transport provided the chosen set of instantaneous observers along the selected geodesic is also parallel along the null curve or possesses special kinematic quantities which further simplify the general transport law \cite{Santana2020}. In the case of a GW in GR, provided a null tetrad decomposition for $\multifieldamp{c}{}{}{(0)\mu\nu}$ is performed, law (iii) also expresses that a notion of ``polarization'' for the GW is parallel transported along $k^\mu$ \cite{Harte2019}.

Finally, subleading-order amplitudes $\multifieldamp{\psi}{}{A}{(p)}$ ($1 \le p \le N$) satisfy a simplified form of Eq.~(\ref{eq:fcg-higher-subleading-order}), but, contrary to the one followed by $\multifieldamp{\psi}{}{A}{(0)}$, including a nonhomogenous term derived from the immediately superleading-order amplitude:
\begin{equation}
	\label{eq:fcg-kinetic-vanishing-friction-subleading-order-transport}
	\multifield{\mathscr{D}}{}{(g)}{}\multifieldamp{\psi}{}{A}{(p+1)} = -\multifield{\mathcal{L}}{}{A}{B} \, \multifieldamp{\psi}{}{B}{(p)} \,, \quad 0 \le p \le N\,,
\end{equation}
where $\multifield{\mathcal{L}}{}{A}{B}$ is determined from the kinetic and mass tensors appearing in Table~\ref{tab:examples-lorentzian-spacetimes} for each tensor field. Eq.~(\ref{eq:fcg-kinetic-vanishing-friction-subleading-order-transport}) gives the \textit{beyond geometrical optics} transport equations for $\multifieldamp{\psi}{}{A}{(p)}$ along the integral curves of the corresponding wave vectors, comprising a set of perturbative equations in which the source term of the transport equation for $\multifieldamp{\psi}{}{A}{(p+1)}$ only depends on the solution of the transport equation for the immediately superleading-order amplitude $\multifieldamp{\psi}{}{A}{(p)}$. These additional transport laws allow one to obtain higher-order corrections to the amplitude associated to the Ansatz of a given field, thus refining the full JWKB amplitude beyond its leading-order component $\multifieldamp{\psi}{}{A}{(0)}$ \cite{Harte2019, Cusin2020}.

\subsection{Coupling and FC-diagonal kinetic tensors: reduced Horndeski theories}
\label{subsec:reduced-horndesk-theories}

After illustraing the generality of our results by considering four examples with FCG-diagonal kinetic tensors and vanishing friction tensors in the previous section, let us turn our attention to a final example in Lorentzian spacetimes, namely, scalar-tensor GWs in a subclass of Horndeski theories of gravity \cite{Horndeski1974, Kobayashi2019} referred to as \textit{reduced Horndeski theories} (RHTs) \cite{Dalang2020, Dalang2021, Lobato2022, Lobato2024}, whose constraints guarantee that scalar-tensor GWs propagate at the speed of light.

Similar to the procedure for GWs in GR, second-order equations of motion can be derived for a GW perturbation pair $(\Phi, H_{\mu\nu})$ by linearizing the corresponding field equations in RHTs. These are coupled equations at zeroth, first and second order derivatives, but as shown in Ref. \cite{Dalang2021}, one can perform a diagonalization procedure to decouple the kinetic terms (second-order derivatives) in the equations for $\Phi$ and $H_{\mu\nu}$, arriving at a new pair ($\multifield{\Psi}{1}{}{} \reqdef \Phi, \multifield{\Psi}{2}{}{\mu\nu} \reqdef \Gamma_{\mu\nu}$) subject to the following equations of motion:
\begin{equation}
	\label{eq:horndeski-equation-diagonal}
	\left[ 
        \begin{array}{cc} 
            \multifield{\mathcal{L}}{11}{}{} & \multifield{\mathcal{L}}{12}{\lambda\sigma}{} \\
            \multifield{\mathcal{L}}{21}{}{\mu\nu} & \multifield{\mathcal{L}}{22}{}{\mu\nu}\tensor{}{^{\lambda\sigma}}
        \end{array}
    \right]
    \left[ 
        \begin{array}{c} 
            \Phi \\
            \Gamma_{\lambda\sigma}
        \end{array}
    \right] = 0 \,,
\end{equation}
where
\begin{equation}
    \label{eq:horndeski-operator}
    \begin{split}
        \left[ 
            \begin{array}{cc} 
                \multifield{\mathcal{L}}{11}{}{} & \multifield{\mathcal{L}}{12}{\lambda\sigma}{} \\
                \multifield{\mathcal{L}}{21}{}{\mu\nu} & \multifield{\mathcal{L}}{22}{}{\mu\nu}\tensor{}{^{\lambda\sigma}}
            \end{array}
        \right]
        \leqdef 
        \left[ 
            \begin{array}{cc} 
                \multifield{\mathcal{K}}{11}{\alpha\beta}{} & 0 \\
                0 & \multifield{\mathcal{K}}{22}{\alpha\beta}{\mu\nu}\tensor{}{^{\lambda\sigma}}
            \end{array}
        \right] 
        \munderbar{\hatnabla}_\alpha \munderbar{\hatnabla}_\beta & + 
        \left[ 
            \begin{array}{cc} 
                \multifield{\mathcal{F}}{11}{\alpha}{} & \multifield{\mathcal{F}}{12}{\alpha\lambda\sigma}{} \\
                \multifield{\mathcal{F}}{21}{\alpha}{\mu\nu} & \multifield{\mathcal{F}}{22}{\alpha}{\mu\nu}\tensor{}{^{\lambda\sigma}}
            \end{array}
        \right]
        \munderbar{\hatnabla}_\alpha + 
        \left[ 
            \begin{array}{cc} 
                \multifield{\mathcal{M}}{11}{}{} & \multifield{\mathcal{M}}{12}{\lambda\sigma}{} \\
                \multifield{\mathcal{M}}{21}{}{\mu\nu} & \multifield{\mathcal{M}}{22}{}{\mu\nu}\tensor{}{^{\lambda\sigma}}
            \end{array}
        \right] \,,
    \end{split}
\end{equation}
with the kinetic, friction, and mass terms determined from the RHT Lagrangian and evaluated in the prescribed background scalar-tensor field $(\munderbar{\varphi}, \munderbar{g}_{\mu\nu})$ [cf. the comment after Eq.~(\ref{eq:field-equation-operator})]. Here we use $\Gamma$ as the kernel letter for the bilinear form without risk of confusion with a general affine connection $\tensor{\Gamma}{^\alpha_{\mu\nu}}$, since parallel transport in Lorentzian spacetimes is given by the Levi-Civita connection of a given metric, which we denote by $\Christ{\alpha}{\mu}{\nu}$ [cf. Eq.~(\ref{eq:affine-connection})]. In recent literature \cite{Dalang2021}, this new pair is interpreted as the true and independent degrees of freedom of scalar-tensor GWs in RHTs, obtained from a transformation that makes $\multifield{\mathcal{K}}{12}{\alpha\beta}{}\tensor{}{^{\lambda\sigma}} = 0 =  \multifield{\mathcal{K}}{21}{\alpha\beta}{\mu\nu}$. The JWKB Ans\"atze in this case read
\begin{equation}
	\label{eq:horndeski-scalar-ansatz}
	\multifield{\Phi}{}{}{}(x, \epsilon) = \left[\,\sum_{p = 0}^N \multifieldamp{\phi}{}{}{(p)}(x) \, \left(\frac{\epsilon}{i}\right)^p\,\right] \, e^{iS_1(x)/\epsilon} \,, \quad N \ge 0 \,,
\end{equation}
and
\begin{equation}
	\label{eq:horndeski-tensor-ansatz}
	\multifield{\Gamma}{}{}{\mu\nu}(x, \epsilon) = \left[\,\sum_{p = 0}^N \multifieldamp{\gamma}{}{}{(p)\mu\nu}(x) \, \left(\frac{\epsilon}{i}\right)^p\,\right] \, e^{iS_2(x)/\epsilon} \,, \quad N \ge 0 \,.
\end{equation}

Now, contrary to the other examples in Lorentzian spacetimes, the kinetic tensors in RHTs are not necessarily FCG-diagonal. However, one can arrive at an FCG-diagonal kinetic tensor for the bilinear form $\Gamma_{\mu\nu}$ by selecting the so-called and always achievable \textit{harmonic gauge} \cite{Dalang2020, Dalang2021}
\begin{equation}
	\label{eq:horndeski-harmonic-gauge}
	\munderbar{\hatnabla}_\nu \Gamma^{\mu\nu} = 0\,,
\end{equation}
which yields
\begin{equation}
	\label{eq:kinetic-horndeski-tensor-harmonic-gauge}
	\multifield{\mathcal{K}}{22}{\alpha\beta}{\mu\nu}\tensor{}{^{\lambda\sigma}} =  \multifield{\overset{(1)}{\varkappa}}{2}{}{} \delta_\mu^\lambda \delta_\nu^\sigma \, \munderbar{g}^{\alpha\beta} \,, \quad \multifield{\overset{(1)}{\varkappa}}{2}{}{} \leqdef -\frac{1}{2} G_4(\munderbar{\varphi}) \reqdef -\frac{1}{2} \munderbar{G}_4 \,,
\end{equation}
where $\munderbar{G}_4$ is one of the nonvanishing terms in the general Lagrangian of RHTs \cite{Dalang2020}. Since $\munderbar{G}_4 \ne 0$, the form of $\multifield{\mathcal{K}}{22}{\alpha\beta}{\mu\nu}\tensor{}{^{\lambda\sigma}}$ implies that the wave vector associated to $S_2(x)$ satisfies [cf. Eq.~(\ref{eq:null-condition})]:
\begin{equation}
	\label{eq:null-condition-horndeski-tensor}
	\munderbar{g}^{\alpha\beta} \, \multifield{k}{2}{}{\alpha} \multifield{k}{2}{}{\beta} = 0 \Rightarrow \multifield{k}{2}{\nu}{} \munderbar{\hatnabla}_\nu \multifield{k}{2}{\mu}{} = 0\,,
\end{equation}
textit{i.e.}, the integral curves of $\multifield{k}{2}{\alpha}{}$ are \textit{null} and \textit{affinely parametrized affine and metric geodesics}. On the other hand, the harmonic gauge does not constrain $\multifield{\mathcal{K}}{11}{\alpha\beta}{}$ in any way. However, since $\Phi$ is a scalar field, \textit{i.e.}, it only possesses a single degree of freedom, $\multifield{\mathcal{K}}{11}{\alpha\beta}{}$ is trivially an FC-diagonal kinetic tensor, meaning that Eq.~(\ref{eq:deviation-nullity-condition}) is valid for $\multifield{k}{1}{\alpha}{}$:
\begin{equation}
	\label{eq:transformed-null-condition-horndeski-scalar}
	\multifield{\mathcal{K}}{11}{\alpha\beta}{} \, \multifield{k}{1}{}{\alpha} \multifield{k}{1}{}{\beta} = \left(\multifield{\overset{(1)}{\varkappa}}{1}{}{} \, \munderbar{g}^{\alpha\beta} + \multifield{\overset{(2)}{\varkappa}}{1}{\alpha\beta}{} \right) \multifield{k}{1}{}{\alpha} \multifield{k}{1}{}{\beta} = 0\,.
\end{equation}

Note that the previous equation does not imply that $\multifield{k}{1}{\alpha}{}$ is null unless $\multifield{\overset{(2)}{\kappa}}{1}{\alpha\beta}{} = 0$. Moreover, these kinetic tensors imply that the transport equations for the leading order amplitudes $\phi_{(0)}$ and $\gamma_{(0)\mu\nu}$ [cf. Eq.~(\ref{eq:fcg-1st-subleading-order})] read \cite{Dalang2021}:
\begin{equation}
	\label{eq:transport-leading-order-horndeski-scalar-simplied}
	\multifield{\widehat{\mathscr{D}}}{\munderbar{1}}{(\mathcal{K})}{} \, \phi_{(0)} + \multifield{\mathcal{F}}{11}{\alpha}{} \, \multifield{k}{1}{}{\alpha} \, \phi_{(0)} + \multifield{\mathcal{F}}{12}{\lambda\sigma}{} \, \multifield{k}{2}{}{\alpha} \, \gamma_{(0)\lambda\sigma} \, e^{i(S_2 - S_1)/\epsilon} = 0 
\end{equation}
and
\begin{equation}
	\label{eq:transport-leading-order-horndeski-tensor-simplied}
	\multifield{\widehat{\mathscr{D}}}{\munderbar{2}}{(g)}{} \, \gamma_{(0)\mu\nu} + \frac{1}{\multifield{\overset{(1)}{\varkappa}}{2}{}{}} \, \multifield{\mathcal{F}}{21}{\alpha}{\mu\nu} \, \multifield{k}{1}{}{\alpha} \, \phi_{(0)} \, e^{i(S_1 - S_2)/\epsilon} + \frac{1}{\multifield{\overset{(1)}{\varkappa}}{2}{}{}} \, \multifield{\mathcal{F}}{22}{\alpha}{\mu\nu}\tensor{}{^{\lambda\sigma}} \, \multifield{k}{2}{}{\alpha} \, \gamma_{(0)\lambda\sigma} = 0\,,
\end{equation}
where $\multifield{\widehat{\mathscr{D}}}{\munderbar{1}}{(\mathcal{K})}{} \leqdef \multifield{\mathcal{K}}{11}{\alpha\beta}{} \, \multifield{\widehat{D}}{\munderbar{1}}{}{\alpha\beta}$. At last, the leading-order relation derived from the harmonic gauge condition for $\Gamma_{\mu\nu}$ implies that $\gamma_{(0)\mu\nu}$ is transverse to $\multifield{k}{2}{\mu}{}$. The above results can be summarized as the \textit{laws of geometrical optics} for scalar-tensor GWs in RHTs:
\begin{enumerate}[(i)]
    \item the scalar pertubation wave vector $\multifield{k}{1}{\mu}{}$ is not null nor (affine and metric) geodesic in general. If, however, $\multifield{\overset{(2)}{\varkappa}}{1}{\alpha\beta}{} = 0$, $\multifield{k}{1}{\mu}{}$ is null and (affine and metric) geodesic, a situation that significantly simplifies the study of scalar-tensor GWs in RHTs \cite{Dalang2020, Dalang2021, Lobato2022, Lobato2024}.
    \item the tensor perturbation wave vector $\multifield{k}{2}{\mu}{}$ is always null and (affine and metric) geodesic.
    \item the leading-order amplitude $\multifieldamp{\phi}{}{}{(0)}$ of the scalar perturbation has no transversality condition with respect to $\multifield{k}{1}{\mu}{}$ given its scalar character.
    \item the leading-order amplitude $\multifieldamp{\gamma}{}{}{(0)\mu\nu}$ of the tensor perturbation  is transverse to $\multifield{k}{2}{\mu}{}$.
    \item $\multifieldamp{\phi}{}{}{(0)}$ evolves according to Eq.~(\ref{eq:transport-leading-order-horndeski-scalar-simplied}).
    \item $\multifieldamp{\gamma}{}{}{(0)\mu\nu}$evolves according Eq.~(\ref{eq:transport-leading-order-horndeski-tensor-simplied}).
\end{enumerate}

Finally, subleading-order amplitudes $\multifieldamp{\psi}{I}{A_I}{(p)}$ ($1 \le p \le N$) satisfy Eqs.~(\ref{eq:fcg-higher-subleading-order}) applied to the current example, but, contrary to the one followed by $\multifieldamp{\psi}{I}{A_I}{(0)}$, including a nonhomogenous term derived from the immediately superleading-order amplitudes:
\begin{equation}
	\label{eq:transport-subleading-order-horndeski-scalar-simplied}
    \begin{split}
	\multifield{\widehat{\mathscr{D}}}{\munderbar{1}}{(\mathcal{K})}{} \phi_{(p+1)} \, e^{iS_1/\epsilon} &+ \multifield{\mathcal{F}}{11}{\alpha}{} \, \multifield{k}{1}{}{\alpha} \, \phi_{(p+1)} \, e^{iS_1/\epsilon} + \\
    &+ \multifield{\mathcal{F}}{12}{\lambda\sigma}{} \, \multifield{k}{2}{}{\alpha} \, \gamma_{(p+1)\lambda\sigma} \, e^{iS_2/\epsilon} = 
    - 
    \left[ 
        \begin{array}{cc} 
            \multifield{\mathcal{L}}{11}{}{} & \multifield{\mathcal{L}}{12}{\lambda\sigma}{} \\
            \multifield{\mathcal{L}}{21}{}{\mu\nu} & \multifield{\mathcal{L}}{22}{}{\mu\nu}\tensor{}{^{\lambda\sigma}}
        \end{array}
    \right] 
    \left[ 
        \begin{array}{c} 
            \phi_{(p)} \, e^{iS_1/\epsilon}\\
            \gamma_{(p)\lambda\sigma} \, e^{iS_2/\epsilon}
        \end{array}
    \right] \,,
    \end{split}
\end{equation}
and
\begin{equation}
	\label{eq:transport-subleading-order-horndeski-tensor-simplied}
    \begin{split}
	\multifield{\widehat{\mathscr{D}}}{\munderbar{2}}{(g)}{} \gamma_{(p+1)\mu\nu} \, e^{iS_2/\epsilon} &+ \frac{1}{\multifield{\overset{(1)}{\varkappa}}{2}{}{}} \, \multifield{\mathcal{F}}{22}{\alpha}{\mu\nu}\tensor{}{^{\lambda\sigma}} \, \multifield{k}{2}{}{\alpha} \, \gamma_{(p+1)\lambda\sigma} \, e^{iS_2/\epsilon} + \\
    &+ \frac{1}{\multifield{\overset{(1)}{\varkappa}}{2}{}{}} \, \multifield{\mathcal{F}}{21}{\alpha}{\mu\nu} \, \multifield{k}{1}{}{\alpha} \, \phi_{(p+1)} \, e^{iS_1/\epsilon} = 
    - 
    \left[ 
        \begin{array}{cc} 
            \multifield{\mathcal{L}}{11}{}{} & \multifield{\mathcal{L}}{12}{\lambda\sigma}{} \\
            \multifield{\mathcal{L}}{21}{}{\mu\nu} & \multifield{\mathcal{L}}{22}{}{\mu\nu}\tensor{}{^{\lambda\sigma}}
        \end{array}
    \right]
    \left[ 
        \begin{array}{c} 
            \phi_{(p)} \, e^{iS_1/\epsilon}\\
            \gamma_{(p)\lambda\sigma} \, e^{iS_2/\epsilon}
        \end{array}
    \right]
    \,,
    \end{split}
\end{equation}

By analogy with the previous examples, Eqs.~(\ref{eq:transport-subleading-order-horndeski-scalar-simplied}) and (\ref{eq:transport-subleading-order-horndeski-tensor-simplied}) give, respectively, the \textit{beyond geometrical optics} transport equations for $\multifieldamp{\phi}{}{}{(p)}$ and $\multifieldamp{\gamma}{}{}{(p)\mu\nu}$ ($1 \le p \le N$) along the integral curves of the corresponding wave vectors. In spite of the fact that $\multifield{k}{1}{\mu}{}$ is not null in general, its associated dispersion operator vanishes [cf. (\ref{eq:vanishing-dispersion-operators})]. As such, all transport equation in this example also comprise a set of perturbative equations, with the source term on the RHS of the transport equations for $\multifieldamp{\phi}{}{}{(p+1)}$ and $\multifieldamp{\gamma}{}{}{(p+1)\mu\nu}$ only depending on the solutions of the transport equations for the immediately superleading-order amplitudes $\multifieldamp{\phi}{}{}{(p)}$ and $\multifieldamp{\gamma}{}{}{(p)\mu\nu}$ ($0 \le p \le N-1$). Notice however, that coupling is generally still present. These additional transport laws allow one to obtain higher-order corrections to the amplitudes associated to the Ans\"atze of the two fields, thus refining the full JWKB amplitudes beyond their leading-order components $\multifieldamp{\psi}{I}{A_I}{(0)}$.

\section{Examples in metric-affine spacetimes}
\label{sec:examples-metric-affine-spacetimes}

In the previous section, we covered five different examples in Lorentzian spacetimes which show how the universal results of Sec. \ref{sec:decomposition-kinetic-operator} can be directly applied to obtain the laws of geometrical optics for a number of tensor fields. Now, the results of Sec. \ref{sec:decomposition-kinetic-operator} were in fact derived for metric-affine spacetimes, which naturally include Lorentzian spacetimes but also modified theories of gravity that include torsion and/or nonmetricity to describe additional degrees of freedom of the gravitational field [cf. App. ~(\ref{append:metric-affine-geometry})]. We now go over two examples in this more general geometrical context in the case of the Faraday tensor.

\subsection[Faraday tensor in MCCAT]{Faraday tensor in spacetimes with a metric-compatible connection and a completely antisymmetric torsion}
\label{subsec:faraday-tensor-completely-antisymmetric-torsion-spacetimes}

In a spacetime with a metric-compatible connection ($\tensor{Q}{_{\mu\nu\alpha}} = 0$) and a completely antisymmetric torsion ($\tensor{T}{_{\mu\nu\alpha}} = -\tensor{T}{_{\nu\mu\alpha}} = \tensor{T}{_{\nu\alpha\mu}} \ne 0$), which we refer to as MCCAT, by analogy with the Lorentzian case, we can use the primitive first-order Maxwell equations as a starting point and recast them in terms of the full affine connection $\tensor{\Gamma}{^\alpha_{\mu\nu}}$:
\begin{equation}
	\label{eq:gauss-ampere-completely-antisymmetric-torsion}
	\hatnabla_\nu F^{\mu\nu} = \nabla_\nu F^{\mu\nu} - \frac{1}{2} \tensor{T}{^\mu_{\alpha\beta}} \, F^{\alpha\beta} = 0\,,
\end{equation}
\begin{equation}
	\label{eq:gauss-faraday-completely-antisymmetric-torsion}
	\partial_{[\lambda} F_{\mu\nu]} = \nabla_{[\lambda} F_{\mu\nu]} - \tensor{T}{^\alpha_{[\mu\nu}} F_{\alpha\lambda]} = 0\,,
\end{equation}
to derive the following wave-like equation of motion for $F_{\mu\nu}$:
\begin{equation}
	\label{eq:faraday-tensor-equation-completely-antisymmetric-torsion}
	\begin{split}
	\square \, F_{\mu\nu} + \delta^\alpha_{[\mu} \, \tensor{T}{_{\nu]}^{\lambda\sigma}} \, \nabla_\alpha F_{\lambda\sigma} + & \left[2\left(\tensor{R}{^\lambda_{[\mu\nu]}^\sigma} + \tensor{R}{^\lambda_{[\mu}} \delta^\sigma_{\nu]} + \nabla_\rho \tensor{T}{^\lambda_{[\mu}^\rho} \delta^\sigma_{\nu]} \right) + \right. \\ & + \nabla_{[\mu} \tensor{T}{_{\nu]}^{\lambda\sigma}} + \left. \nabla^\sigma \tensor{T}{^\lambda_{\mu\nu}} - (1/2) \tensor{T}{^\rho_{\mu\nu}} \tensor{T}{_\rho^{\lambda\sigma}}\right] F_{\lambda\sigma} = 0\,,
	\end{split}
\end{equation}
where we recall that the original Maxwell equations need to be further enforced on the solutions of the previous equations to remove nonphysical solutions. A quick inspection shows that the kinetic, friction and mass tensors are, respectively, given by:
\begin{equation}
	\label{eq:faraday-tensor-operators-completely-antisymmetric-torsion-kinetic-friction}
	\multifield{\mathcal{K}}{}{\alpha\beta}{\mu\nu}\tensor{}{^{\lambda\sigma}} = \delta_{\mu}^\lambda \, \delta_{\nu}^\sigma \, g^{\alpha\beta}\,, 
	\quad 
	\multifield{\mathcal{F}}{}{\alpha}{\mu\nu}\tensor{}{^{\lambda\sigma}} = \delta^\alpha_{[\mu} \, \tensor{T}{_{\nu]}^{\lambda\sigma}}\,, 
\end{equation}
\begin{equation}
	\label{eq:faraday-tensor-operators-completely-antisymmetric-torsion-mass}
	\multifield{\mathcal{M}}{}{}{\mu\nu}\tensor{}{^{\lambda\sigma}} = 2\left(\tensor{R}{^\lambda_{[\mu\nu]}^\sigma} + \tensor{R}{^\lambda_{[\mu}} \delta^\sigma_{\nu]} + \nabla_\rho \tensor{T}{^\lambda_{[\mu}^\rho} \delta^\sigma_{\nu]} \right) + \nabla_{[\mu} \tensor{T}{_{\nu]}^{\lambda\sigma}} + \nabla^\sigma \tensor{T}{^\lambda_{\mu\nu}} - (1/2) \tensor{T}{^\rho_{\mu\nu}} \tensor{T}{_\rho^{\lambda\sigma}} \,.
\end{equation}

The Ansatz is equal to the one for the Lorentzian case, namely, Eq.~(\ref{eq:general-ansatz-no-coupling}) applied to $F_{\mu\nu}$. Now, even with a more complicated equation of motion, the kinetic tensor here is also FCG-diagonal, such that the wave vector here as well is \textit{null} [cf. Eq.~(\ref{eq:null-condition})] and \textit{affinely parametrized affine and metric geodesic} [cf. the comment after Eq.~(\ref{eq:wave-vector-transport})] \cite{Santana2017}. This condition on the kinetic tensor and the nonzero friction tensor for MCCAT lead to the following transport equation for the leading-order amplitude $f_{(0)\mu\nu}$ [cf. Eq.~(\ref{eq:fcg-1st-subleading-order})]:
\begin{equation}
	\label{eq:faraday-tensor-completely-antisymmetric-torsion-transport-leading-order}
	\multifield{\mathscr{D}}{}{(g)}{} \multifieldamp{f}{}{}{(0)\mu\nu} + \multifield{k}{}{}{[\mu} \tensor{T}{_{\nu]}^{\lambda\sigma}} \, \multifieldamp{f}{}{}{(0)\lambda\sigma} = 0\,,
\end{equation}
where
\begin{equation}
	\label{eq:g-scalar-transport-operator-completely-antisymmetric-torsion}
	\multifield{\mathscr{D}}{}{(g)}{} = 2 \, \multifield{k}{}{\alpha}{} \nabla_\alpha + \nabla_\alpha \multifield{k}{}{\alpha}{}\,.
\end{equation}

Now, the gauge conditions to be imposed on the solutions above are given by Eqs.~(\ref{eq:gauss-ampere-completely-antisymmetric-torsion}) and (\ref{eq:gauss-faraday-completely-antisymmetric-torsion}), implying that $\multifieldamp{f}{}{}{(0)\mu\nu}$ is transverse to the wave vector $k^\mu$. Therefore, the \textit{laws of geometrical optics} for the Faraday tensor herein read:
\begin{enumerate}[(i)]
    \item the wave vector $k^\mu$ is null and (affine and metric) geodesic, 
    \item the leading-order amplitude $\multifieldamp{f}{}{}{(0)\mu\nu}$ is transverse to $k^\mu$, and
    \item $\multifieldamp{f}{}{}{(0)\mu\nu}$ is transported according to Eq.~(\ref{eq:faraday-tensor-completely-antisymmetric-torsion-transport-leading-order}), which in the case of MCCAT has an additional friction tensor.
\end{enumerate}

Finally, subleading-order amplitudes $\multifieldamp{f}{}{}{(p)\mu\nu}$ ($1 \le p \le N$) satisfy Eq.~(\ref{eq:fcg-higher-subleading-order}) for the MCCAT case:
\begin{equation}
	\label{eq:faraday-tensor-completely-antisymmetric-torsion-transport-subleading-order}
	\multifield{\mathscr{D}}{}{(g)}{} \multifieldamp{f}{}{}{(p+1)\mu\nu} + \multifield{k}{}{}{[\mu} \tensor{T}{_{\nu]}^{\lambda\sigma}} \, \multifieldamp{f}{}{}{(p+1)\lambda\sigma} = -\multifield{\mathcal{L}}{}{}{\mu\nu}\tensor{}{^{\lambda\sigma}} \, \multifieldamp{f}{}{}{(p)\lambda\sigma}\,, \quad (0 \ge p \ge N)
\end{equation}
with
\begin{equation}
    \label{eq:eom-operator-faraday-completely-antisymmetric-torsion}
    \begin{split}
        \multifield{\mathcal{L}}{}{}{\mu\nu}\tensor{}{^{\lambda\sigma}} = \delta_{\mu}^\lambda \, \delta_{\nu}^\sigma \, \square + \delta^\alpha_{[\mu} \, \tensor{T}{_{\nu]}^{\lambda\sigma}} \, \nabla_\alpha & + 2\left(\tensor{R}{^\lambda_{[\mu\nu]}^\sigma} + \tensor{R}{^\lambda_{[\mu}} \delta^\sigma_{\nu]} + \nabla_\rho \tensor{T}{^\lambda_{[\mu}^\rho} \delta^\sigma_{\nu]} \right) + \\ & + \nabla_{[\mu} \tensor{T}{_{\nu]}^{\lambda\sigma}} + \nabla^\sigma \tensor{T}{^\lambda_{\mu\nu}} - (1/2) \tensor{T}{^\rho_{\mu\nu}} \tensor{T}{_\rho^{\lambda\sigma}}\,.
    \end{split}
\end{equation}

Eq.~(\ref{eq:faraday-tensor-completely-antisymmetric-torsion-transport-leading-order}) shows that $\multifieldamp{f}{}{}{(0)\mu\nu}$ evolves independently of the higher-order amplitudes $\multifieldamp{f}{}{}{(p)\mu\nu}$ ($1 \le p \le N$), similar to all the Lorentzian examples we considered in Sec. \ref{sec:examples-lorentzian-spacetimes}. Additionally, Eqs.~(\ref{eq:faraday-tensor-completely-antisymmetric-torsion-transport-subleading-order}) are \textit{beyond geometrical optics} transport equations for $\multifieldamp{f}{}{}{(p)\mu\nu}$ ($1 \le p \le N$) along the integral curves of $k^\mu$. These also comprise a set of perturbative equations, allowing one to obtain higher-order corrections to the amplitude of the Ansatz of $F_{\mu\nu}$, thus refining the full JWKB amplitude beyond its leading-order component $\multifieldamp{f}{}{}{(0)\mu\nu}$.

\subsection{Faraday tensor in Weyl spacetimes}
\label{subsec:faraday-tensor-weyl-spacetimes}

In a spacetime with a symmetric connection ($\tensor{T}{^\alpha_{\mu\nu}} = 0$) and a Weyl nonmetricity ($\tensor{Q}{_{\mu\nu\alpha}} = g_{\mu\nu} \, \zeta_\alpha$, where $\zeta_\alpha$ is an arbitrary real 1-form), we continue to use Maxwell equations as a starting point and recast them in terms of the full affine connection $\tensor{\Gamma}{^\alpha_{\mu\nu}}$
\begin{equation}
    \hatnabla_\nu F^{\mu\nu} = \nabla_\nu F^{\mu\nu} + 2\zeta_\nu \, F^{\mu\nu} = 0\,,    
\end{equation}
\begin{equation}
    \partial_{[\lambda} F_{\mu\nu]} = \nabla_{[\lambda} F_{\mu\nu]} = 0\,.
\end{equation}

These imply that the Faraday tensor satisfies the following wave-like equation of motion:
\begin{equation}
    \label{eq:faraday-tensor-equation-weyl}
    \square \, F_{\mu\nu} + 2 \zeta_{[\mu} \, g^{\alpha\sigma} \delta^\lambda_{\nu]} \, \nabla_\alpha F_{\lambda\sigma} + 2\left(\tensor{R}{^\lambda_{[\mu\nu]}^\sigma} + \tensor{R}{^\lambda_{\rho[\mu}^\rho} \delta^\sigma_{\nu]} \right) F_{\lambda\sigma} = 0\,.
\end{equation}

A quick inspection shows that the kinetic, friction and mass tensors in this case are, respectively, given by:
\begin{equation}
    \label{eq:faraday-tensor-operators-weyl-kinetic-friction}
    \multifield{\mathcal{K}}{}{\alpha\beta}{\mu\nu}\tensor{}{^{\lambda\sigma}} = \delta_{\mu}^\lambda \, \delta_{\nu}^\sigma \, g^{\alpha\beta}\,, 
    \quad 
    \multifield{\mathcal{F}}{}{\alpha}{\mu\nu}\tensor{}{^{\lambda\sigma}} = 2 \zeta_{[\mu} \, g^{\alpha\sigma} \delta^\lambda_{\nu]}\,, 
\end{equation}
\begin{equation}
    \label{eq:faraday-tensor-operators-weyl-mass}
    \multifield{\mathcal{M}}{}{}{\mu\nu}\tensor{}{^{\lambda\sigma}} = 2\left(\tensor{R}{^\lambda_{[\mu\nu]}^\sigma} + \tensor{R}{^\lambda_{\rho[\mu}^\rho} \delta^\sigma_{\nu]}\right) \,.
\end{equation}

The proposed Ansatz is equal to the one for the Lorentzian case and the first metric-affine geometry we have considered. Here as well, the kinetic tensor is FCG-diagonal, such that the wave vector is also \textit{null} [cf. Eq.~(\ref{eq:null-condition})]. However, contrary to MCCAT, for a Weyl connection, Eq.~(\ref{eq:wave-vector-transport}) implies that integral curves of $\multifield{k}{}{}{\alpha}$ are affinely parametrized metric geodesics but nonaffinely parametrized autoparallels [cf. again the comment after Eq.~(\ref{eq:wave-vector-transport})]. Finally, the leading-order of Eq.~(\ref{eq:fcg-1st-subleading-order}), given that we also have a nonzero friction, yields
\begin{equation}
    \label{eq:faraday-tensor-weyl-transport-leading-order}
    \multifield{\mathscr{D}}{}{(g)}{} \multifieldamp{f}{}{}{(0)\mu\nu} + 2 \zeta_{[\mu} \, k^\sigma \delta^\lambda_{\nu]} \, \multifieldamp{f}{}{}{(0)\lambda\sigma} = 0\,,
\end{equation}
whereas Eqs.(\ref{eq:fcg-higher-subleading-order}) leads to the transport equations of subleading-order amplitudes:
\begin{equation}
    \label{eq:faraday-tensor-weyl-transport-subleading-order}
    \multifield{\mathscr{D}}{}{(g)}{} \multifieldamp{f}{}{}{(p+1)\mu\nu} + 2 \zeta_{[\mu} \, k^\sigma \delta^\lambda_{\nu]} \, \multifieldamp{f}{}{}{(p+1)\lambda\sigma} = -\multifield{\mathcal{L}}{}{}{\mu\nu}\tensor{}{^{\lambda\sigma}} \, \multifieldamp{f}{}{}{(p)\lambda\sigma}\,, \quad (0 \ge p \ge N)
\end{equation}
with
\begin{equation}
    \label{eq:g-scalar-transport-operator-weyl}
    \multifield{\mathscr{D}}{}{(g)}{} = 2 \, \multifield{k}{}{\alpha}{} \nabla_\alpha + \nabla_\alpha \multifield{k}{}{\alpha}{} + \multifield{k}{}{\alpha}{} \zeta_\alpha \,,
\end{equation}
and
\begin{equation}
    \label{eq:eom-operator-faraday-weyl}
    \multifield{\mathcal{L}}{}{}{\mu\nu}\tensor{}{^{\lambda\sigma}} = \delta_{\mu}^\lambda \, \delta_{\nu}^\sigma \, \square + 2 \zeta_{[\mu} \, g^{\alpha\sigma} \delta^\lambda_{\nu]} \, \nabla_\alpha + 2\left(\tensor{R}{^\lambda_{[\mu\nu]}^\sigma} + \tensor{R}{^\lambda_{\rho[\mu}^\rho} \delta^\sigma_{\nu]} \right)\,.
\end{equation}

Finally, by imposing Maxwell equations as gauge conditions on the above solutions shows that $\multifieldamp{f}{}{}{(0)\mu\nu}$ is transverse to the wave vector. Similar to the MCCAT case, the above constraints imply that Eq.~(\ref{eq:faraday-tensor-weyl-transport-leading-order}) simplifies to
\begin{equation}
    \label{eq:faraday-tensor-weyl-transport-leading-order-simplified}
    \multifield{\mathscr{D}}{}{(g)}{} \multifieldamp{f}{}{}{(0)\mu\nu} = 0\,,
\end{equation}
where the above differential operator resembles the one appearing in the Lorentzian examples, but, in Weyl spacetimes, the 1-form $\zeta_\alpha$ modifies the simpler $\multifield{\widehat{\mathscr{D}}}{}{(g)}{}$.

Accordingly, the \textit{laws of geometrical optics} for the Faraday tensor read:
\begin{enumerate}[(i)]
    \item the wave vector $k^\mu$ is null and (affine and metric) geodesic, 
    \item the leading-order amplitude $\multifieldamp{f}{}{}{(0)\mu\nu}$ is transverse to $k^\mu$, and
    \item $\multifieldamp{f}{}{}{(0)\mu\nu}$ is transported according to Eq.~(\ref{eq:faraday-tensor-weyl-transport-leading-order}).
\end{enumerate}

Eq.(\ref{eq:faraday-tensor-weyl-transport-leading-order}) shows that $\multifieldamp{f}{}{}{(0)\mu\nu}$ evolves independently of the higher-order amplitudes $\multifieldamp{f}{}{}{(p)\mu\nu}$ ($1 \le p \le N$). In turn, Eqs.~(\ref{eq:faraday-tensor-weyl-transport-subleading-order}) are \textit{beyond geometrical optics} transport equations for $\multifieldamp{f}{}{}{(p)\mu\nu}$ ($1 \le p \le N$) along the integral curves of $k^\mu$. These also comprise a set of perturbative equations, allowing one to obtain higher-order corrections to the amplitude of the Ansatz of $F_{\mu\nu}$, thus refining the full JWKB amplitude beyond its leading-order component $\multifieldamp{f}{}{}{(0)\mu\nu}$.

\begin{figure}[h!]
    \centering
    \includegraphics[scale = 0.6]{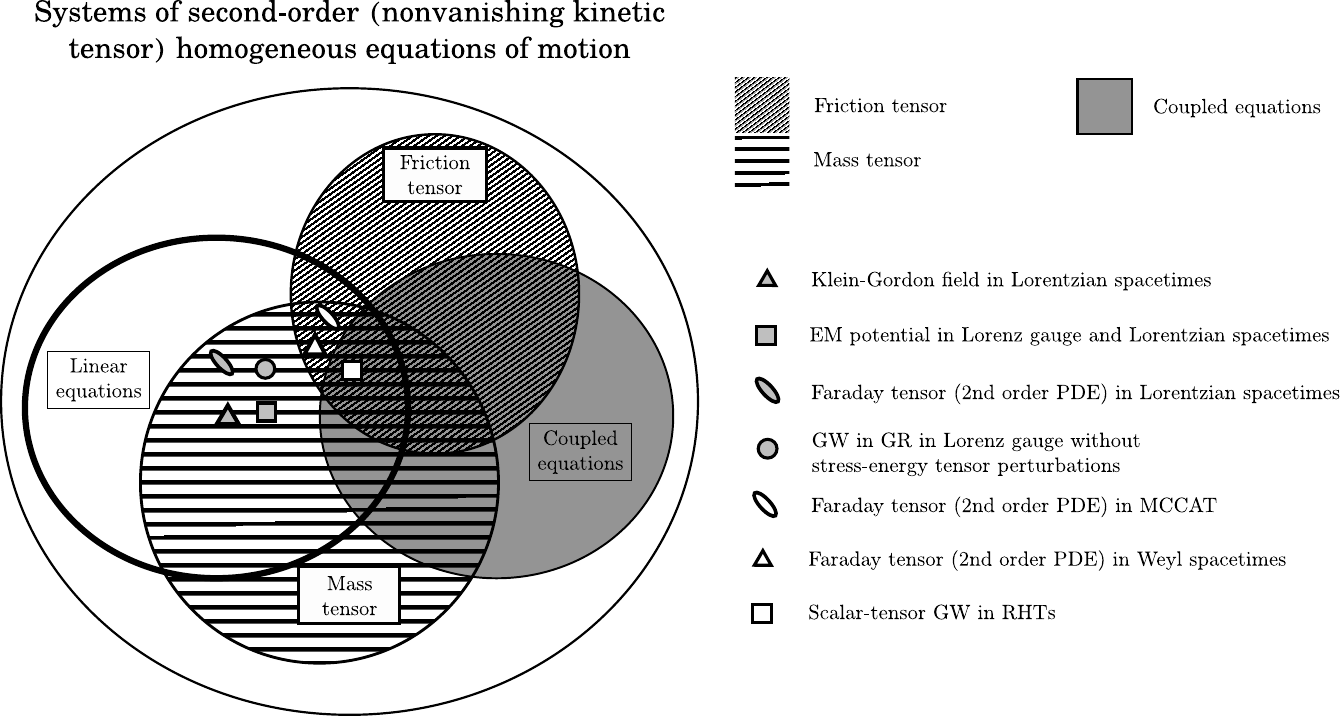}
    \caption{Venn diagram representing the different possible systems of second-order (nonvanishing kinetic tensor) homogeneous equations of motion and the presence or absence of friction and mass tensors. All examples in Secs.~\ref{sec:examples-lorentzian-spacetimes} and \ref{sec:examples-metric-affine-spacetimes} appear in relevant intersections among the sets considered.}
    \label{fig:venn-diagram}
\end{figure}

\section{Discussion}
\label{sec:discussion}

In this work, we have developed a unified framework for describing linear waves associated to tensor fields of arbitrary rank obeying coupled homogeneous linear wave-like second-order PDEs. Even though Eqs.~(\ref{eq:leading-order-general}) through (\ref{eq:higher-subleading-order-general}) are quite straightforwardly derived from inserting the Ans\"{a}tze (\ref{eq:rewritten-form-ansatze}) into Eqs.~(\ref{eq:field-equation}) and demanding that the latter are valid for each and every order of $\epsilon$, those relations are the core results leading to the well-known relations of GO and BGO appearing in the literature. 

More specifically, GO is derived from the leading ($p=0$) and first subleading-orders ($p=1$) of those equations, which give, respectively, definite constraints among the vector fields $\multifield{q}{I}{}{\alpha}$ and transport equations for every leading-order amplitude $\multifieldamp{\psi}{I}{A_I}{(0)}$, whereas BGO relates to the transport equations for the ($p \ge 1$)-order amplitudes. Furthermore, the GO results only depend on the kinetic and friction tensors appearing in Eqs.(\ref{eq:field-equation}). This is precisely why we considered the decomposition of a generic kinetic tensor in Sec.~ \ref{sec:decomposition-kinetic-operator}, which allowed us to establish important results directly dependent on its form, namely, F-diagonality, FG-diagonality, FC-diagonality, and FCG-diagonality. 

As we have seen, provided that $\multifieldamp{\psi}{I}{A_I}{(0)}$ vanishes at most in isolated points, either FG-diagonality or FCG-diagonality for the kinetic tensor lead to $\multifield{q}{I}{}{\alpha}$ being null, which, together with $\multifield{q}{I}{}{\alpha}$ being a gradient, yields Eq.~(\ref{eq:wave-vector-transport}) as its corresponding transport law. This is one of the core results derived from the unified framework we present herein, showing that specific forms for the kinetic tensor---which in the language of characteristic surfaces and bi-characteristic curves \cite{Bona2011} correspond to the so-called \textit{principal part} of the associated equations of motion---directly lead to a generalization of one of the key results of traditional GO, namely, that rays are \textit{null geodesics} \cite{Ehlers1967, Ehlers2022, Ellis2012, Perlick2000, Misner1973}. 

This result applies to all examples in Lorentzian spacetimes we have considered in Sec.~ \ref{sec:examples-lorentzian-spacetimes}, with the exception of RHTs, in which the perturbation of the scalar field does not necessarily possess either an FG- or an FCG-diagonal kinetic tensor. Furthemore, given a more general affine connection, Eq.~(\ref{eq:wave-vector-transport}) shows that the null curves of a given physical field continue to be extremal geodesics, but no longer need to be autoparallels \cite{Santana2017}. And even though this is the case, in Sec.~ \ref{sec:examples-metric-affine-spacetimes} we gave two explicit nontrivial examples, MCCAT and Weyl spacetimes, where the null curves continue to be autoparallels (nonaffinely parametrized in the latter case).

Hence, contrary to Lorentzian geometry, where traditional GO gives the pleasant result that null curves are geodesics---which demonstrates in an alternative manner Einstein's original hypothesis in his formulation of GR \cite{Einstein1916}---, a similar formalism applied to metric-affine gravity gives a perhaps not-so-pleasant but still consistent outcome. Nonetheless, even in the more general context of metric-affine spacetimes, provided we still assume fields satisfy linear equations of motion, the first law of geometrical optics remains independent of any amplitude $\multifieldamp{\psi}{I}{A_I}{(p)}$, meaning the geometry of null rays can be studied irrespective of how the amplitude of the corresponding field evolves through spacetime. This is precisely what allows us to properly study the geometry of null rays and generalize, for example, the usual distance reciprocity relation in the context of an arbitrary metric-affine spacetime \cite{Santana2017}. 

While beyond the scope of this presentation, provided the associated kinetic tensor has an FCG-diagonal form, the GO transport equation for an otherwise arbitrary $\multifieldamp{\psi}{I}{A_I}{(p)}$ [cf. Eq.~(\ref{eq:fcg-1st-subleading-order})] can be used to derive balance equations for a phenomenological brightness-related quantity, from which a notion of particle number arises by making use of the de Broglie hypothesis. This would allow us to obtain a first principles approach to study the relativistic thermodynamic regime of such systems \cite{Calvao1992} or to deal with kinetic treatments for gravitationally induced particle production \cite{Lima2014}. In a future work, we shall explore such phenomenological equations to derive a generalized version of the usual distance duality relation in a generic metric-affine spacetime. In particular, such generalization could be contrasted to observational data in order to constrain additional degrees of freedom in the gravitational field and/or exotic types of matter and radiation inhabiting the Universe.

\section{Acknowledgments}
JCL thanks Brazilian funding agency CAPES for PhD scholarship 88887.492685/2020-00. RRRR acknowledges CNPq (grant no. 309868/2021-1). The authors would like to thank the anonymous referee for the meticulous review of the manuscript and constructive feedback, which contributed to the improvement of this work.

\appendix

\section{Metric-affine geometry}
\label{append:metric-affine-geometry}

A metric-affine spacetime is defined as a triple $(\mathscr{M}, g_{\alpha\beta}, \tensor{\Gamma}{^\mu_{\alpha\beta}})$, $\mathscr{M}$ is a 4-dimensional smooth manifold, $g_{\alpha\beta}$, a Lorentzian metric with signature $+2$, and the affine connection $\tensor{\Gamma}{^\mu_{\alpha\beta}}$ is, in general, independent of the metric tensor. In particular, the latter is such that there are additional tensors describing the curvature of spacetime, namely, the \textit{torsion}
\begin{equation}
    \label{eq:torsion}
    \tensor{T}{^\mu_{\alpha\beta}} \leqdef \tensor{\Gamma}{^\mu_{\alpha\beta}} - \tensor{\Gamma}{^\mu_{\beta\alpha}} \,,
\end{equation}
and the \textit{nonmetricity}
\begin{equation}
    \label{eq:non-metricity}
    \tensor{Q}{_{\alpha\beta\mu}} \leqdef \nabla_\mu g_{\alpha\beta}\,.
\end{equation}
With these definitions, we are able write the generic connection components in any coordinate chart in the form 
\begin{equation}
    \label{eq:affine-connection}
    \tensor{\Gamma}{^\mu_{\alpha\beta}} = \Christ{\mu}{\alpha}{\beta} + \tensor{K}{^\mu_{\alpha\beta}} + \tensor{D}{^\mu_{\alpha\beta}}\,,
\end{equation}
where
\begin{equation}
    \label{eq:levi-civita-connection}
    \Christ{\mu}{\alpha}{\beta} \leqdef \frac{1}{2} g^{\mu\nu} \left( \partial_\beta g_{\nu\alpha} + \partial_\alpha g_{\beta\nu} - \partial_\nu g_{\alpha\beta} \right)\,,
\end{equation}
\begin{equation}
    \label{eq:contorsion}
    \tensor{K}{^\mu_{\alpha\beta}} \leqdef \frac{1}{2} \left(\tensor{T}{_{\alpha\beta}^\mu} + \tensor{T}{_{\beta\alpha}^\mu} + \tensor{T}{^\mu_{\alpha\beta}} \right)\,,
\end{equation}
and
\begin{equation}
    \label{eq:deflection}
    \tensor{D}{^\mu_{\alpha\beta}} \leqdef \frac{1}{2} \, \left(\tensor{Q}{_{\alpha\beta}^\mu} - \tensor{Q}{^\mu_{\alpha\beta}} - \tensor{Q}{^\mu_{\beta\alpha}} \right)\,.
\end{equation}

The first term in Eq.~(\ref{eq:levi-civita-connection}) is the usual \textit{Levi-Civita connection} associated to the metric $g_{\alpha\beta}$, whereas Eqs.~(\ref{eq:contorsion}) and (\ref{eq:deflection}) define, respectively, the \textit{contorsion} and the \textit{deflection} tensors \cite{Santana2017}. In the presence of torsion, the Ricci identity reads:
\begin{eqnarray}
    \label{eq:ricci-identity-metric-affine}
    (\nabla_\nu \nabla_\mu - \nabla_\mu \nabla_\nu ) \, \tensor{\Psi}{^{\alpha_1 \cdots \alpha_r}_{\beta_1 \cdots \beta_s}} &= \sum_{i = 1}^r \tensor{R}{^{\alpha_i}_{\lambda\mu\nu}} \, \tensor{\Psi}{^{\alpha_1 \cdots \lambda \cdots \alpha_r}_{\beta_1 \cdots \beta_s}} + \\
    & - \sum_{i = 1}^s \tensor{R}{^{\lambda}_{\beta_i\mu\nu}} \, \tensor{\Psi}{^{\alpha_1 \cdots \alpha_r}_{\beta_1 \cdots \lambda \cdots \beta_s}} + \\
    & - \tensor{T}{^\lambda_{\mu\nu}} \, \nabla_\lambda \, \tensor{\Psi}{^{\alpha_1 \cdots \alpha_r}_{\beta_1 \cdots \beta_s}} \,,
\end{eqnarray}
where we have defined the \textit{Riemann tensor} of the generic connection $\tensor{\Gamma}{^{\mu}_{\alpha\beta}}$:
\begin{equation}
    \label{eq:riemann-tensor}
    \tensor{R}{^{\alpha}_{\beta\mu\nu}} \leqdef \partial_\nu \tensor{\Gamma}{^{\alpha}_{\beta\mu}} - \partial_\mu \tensor{\Gamma}{^{\alpha}_{\beta\nu}} + \tensor{\Gamma}{^{\alpha}_{\lambda\nu}} \, \tensor{\Gamma}{^{\lambda}_{\beta\mu}} - \tensor{\Gamma}{^{\alpha}_{\lambda\mu}} \, \tensor{\Gamma}{^{\lambda}_{\beta\nu}}s \,.
\end{equation}

\bibliographystyle{unsrt}


\begin{thebibliography}{37}
	
	\bibitem{Schneider1992}
	P.~Schneider, J.~Ehlers, and E.~E. Falco.
	\newblock {\em Gravitational {L}enses}.
	\newblock Springer-Verlag, Berlin, 1992.
	
	\bibitem{Abbott2016}
	B.~P. Abbott.
	\newblock {GW}150914: The {A}dvanced {LIGO} detectors in the era of first
	discoveries.
	\newblock {\em Phys. Rev. Lett.}, 116, 2016.
	
	\bibitem{Hadamard1923}
	J.~Hadamard.
	\newblock {\em Lectures on Cauchy's problem in linear partial differential
		equations}.
	\newblock Yale University Press, 1923.
	
	\bibitem{Bremmer1951}
	H.~Bremmer.
	\newblock The jumps of discontinuous solutions of the wave equation.
	\newblock {\em Comm. Pure Appl. Math.}, 4:419, 1951.
	
	\bibitem{Kline1965}
	M.~Kline and I.~W. Kay.
	\newblock {\em Electromagnetic {T}heory and {G}eometrical {O}ptics}.
	\newblock John Wiley \& Sons, New York, 1965.
	
	\bibitem{Friedlander1975}
	F.~G. Friedlander.
	\newblock {\em The {W}ave {E}quation on a {C}urved {S}pace-{T}ime}.
	\newblock Cambridge University Press, Cambridge, 1975.
	
	\bibitem{Courant1989}
	R.~Courant and D.~Hilbert.
	\newblock {\em Methods of {M}athematical {P}hysics}.
	\newblock Wiley-VCH, Berlin, second edition, 1989.
	
	\bibitem{Born1999}
	M.~Born.
	\newblock {\em Principles of optics}.
	\newblock Cambridge University Press, 1999.
	
	\bibitem{Bona2011}
	A.~B\'{o}na and M.~A. Slawinski.
	\newblock {\em Wavefronts and {R}ays as {C}haracteristics and {A}symptotics}.
	\newblock World Scientific, Singapore, 2011.
	
	\bibitem{Reall2014}
	H.~S. Reall, N.~Tanahashi, and B.~Way.
	\newblock Causality and hyperbolicity of {L}ovelock theories.
	\newblock {\em Class. Quantum Grav.}, 31(20):205005, October 2014.
	
	\bibitem{Tanahashi2017}
	N.~Tanahashi and S.~Ohashi.
	\newblock Wave propagation and shock formation in the most general
	scalar–tensor theories.
	\newblock {\em Class. Quantum Grav.}, 34(21):215003, September 2017.
	
	\bibitem{Horsley2011}
	S.~A.~R. Horsley.
	\newblock Transformation optics, isotropic chiral media and non-{R}iemannian
	geometry.
	\newblock {\em New J. Phys.}, 13:053053, 2011.
	
	\bibitem{Ehlers1967}
	J.~Ehlers.
	\newblock Zum \"{U}bergang von der {W}ellenoptik zur geometrischen {O}ptik in
	der allgemeinen {R}elativit\"{a}tstheorie.
	\newblock {\em Z. Naturforsch.}, 22a:1328, 1967.
	
	\bibitem{Misner1973}
	C.~W. Misner, K.~S. Thorne, and J.~A. Wheeler.
	\newblock {\em Gravitation}.
	\newblock Freeman, San Francisco, 1973.
	
	\bibitem{Anile1989}
	A.~M. Anile.
	\newblock {\em Relativistic {F}luids and {M}agneto-fluids with {A}pplications
		in {A}strophysics and {P}lasma {P}hysics}.
	\newblock Cambridge University Press, Cambridge, UK, 1989.
	
	\bibitem{Perlick2000}
	V.~Perlick.
	\newblock {\em Ray {O}ptics, {F}ermat's {P}rinciple, and {A}pplications to
		{G}eneral {R}elativity}.
	\newblock Springer-Verlag, Berlin, 2000.
	
	\bibitem{ChoquetBruhat2009}
	Y.~Choquet-Bruhat.
	\newblock {\em General relativity and the {E}instein equations}.
	\newblock Oxford University Press, 2009.
	
	\bibitem{Ellis2012}
	G.~F.~R. Ellis, R.~Maartens, and M.~A.~H. MacCallum.
	\newblock {\em Relativistic cosmology}.
	\newblock Cambridge University Press, Cambridge, United Kingdom, 2012.
	
	\bibitem{Harte2019}
	A.~I. Harte.
	\newblock Gravitational lensing beyond geometric optics: I. {F}ormalism and
	observables.
	\newblock {\em Gen. Relativ. Gravit.}, 51, 2019.
	
	\bibitem{Cusin2019}
	G.~Cusin, R.~Durrer, and P.~G. Ferreira.
	\newblock Polarization of a stochastic gravitational wave background through
	diffusion by massive structures.
	\newblock {\em Phys. Rev. D}, 99, 2019.
	
	\bibitem{Kravtsov1990}
	Y.~Kravtsov and Y.~Orlov.
	\newblock {\em {G}eometrical {O}ptics of {I}nhomogeneous {M}edia}.
	\newblock Springer, 1990.
	
	\bibitem{Cusin2020}
	G.~Cusin and M.~Lagos.
	\newblock Gravitational wave propagation beyond geometric optics.
	\newblock {\em Phys. Rev. D}, 101(4):044041, 2020.
	
	\bibitem{Santana2017}
	L.~T. Santana, M.~O. Calv\~{a}o, R.~R.~R. Reis, and B.~B. Siffert.
	\newblock How does light move in a generic metric-affine background?
	\newblock {\em Phys. Rev. D}, 95:061501(R), 2017.
	
	\bibitem{Dalang2020}
	C.~Dalang, P.~Fleury, and L.~Lombriser.
	\newblock Horndeski gravity and standard sirens.
	\newblock {\em Phys. Rev. D}, 102:044036, 2020.
	
	\bibitem{Dalang2021}
	C.~Dalang, P.~Fleury, and L.~Lombriser.
	\newblock Scalar and tensor gravitational waves.
	\newblock {\em Phys. Rev. D}, 103:064075, 2021.
	
	\bibitem{Lobato2022}
	J.~C. Lobato, I.~S. Matos, M.~O. Calv{\~{a}}o, and I.~Waga.
	\newblock Gravitational wave stochastic background in reduced {H}orndeski
	theories.
	\newblock {\em Phys. Rev. D}, 106:104048, 2022.
	
	\bibitem{Lobato2024}
	J.~C. Lobato and M.~O. Calv\~ao.
	\newblock Gravitational wave energy-momentum tensor in reduced {H}orndeski
	theories.
	\newblock {\em Phys. Rev. D}, 109:044004, 2024.
	
	\bibitem{Santana2020}
	L.~T. Santana, J.~C. Lobato, I.~S. Matos, M.~O. Calv\~{a}o, and R.~R.~R. Reis.
	\newblock Evolution of the electric field along null rays for arbitrary
	observers and spacetimes.
	\newblock {\em Phys. Rev. D}, 101:081501(R), 2020.
	
	\bibitem{Sachs1977}
	R.~K. Sachs and H.~H. Wu.
	\newblock {\em General {R}elativity for {M}athematicians}.
	\newblock Springer-Verlag, New York, USA, 1977.
	
	\bibitem{Lobato2021}
	J.~C. Lobato, I.~S. Matos, L.~T. Santana, R.~R.~R. Reis, and M.~O.
	Calv{\~{a}}o.
	\newblock Influence of gravitational waves upon light in the {M}inkowski
	background: {F}rom null geodesics to interferometry.
	\newblock {\em Phys. Rev. D}, 104:024024, 2021.
	
	\bibitem{Harte2013}
	A.~I. Harte.
	\newblock Strong lensing, plane gravitational waves and transient flashes.
	\newblock {\em Class. Quantum Grav.}, 30:075011, 2013.
	
	\bibitem{Ehlers2022}
	J.~Ehlers.
	\newblock Republication of: {O}n the transition from wave optics to geometric
	optics in general relativity.
	\newblock {\em Gen. Relativ. Gravit.}, 54, 2022.
	
	\bibitem{Horndeski1974}
	G.~W. Horndeski.
	\newblock Second-order scalar-tensor field equations in a four-dimensional
	space.
	\newblock {\em Int. J. Theor. Phys.}, 10:363, 1974.
	
	\bibitem{Kobayashi2019}
	T.~Kobayashi.
	\newblock Horndeski theory and beyond: a review.
	\newblock {\em Reports on Progress in Physics}, 82:086901, 2019.
	
	\bibitem{Einstein1916}
	A.~Einstein.
	\newblock Die {G}rundlage der allgemeinen {R}elativitätstheorie.
	\newblock {\em Ann. {P}hys.}, 354:769, 1916.
	
	\bibitem{Calvao1992}
	M.~O. Calv{\~a}o, J.~A.~S. Lima, and I.~Waga.
	\newblock On the thermodynamics of matter creation in cosmology.
	\newblock {\em Phys. Lett. A}, 162:223, 1992.
	
	\bibitem{Lima2014}
	J.~A.~S. Lima and I.~Baranov.
	\newblock Gravitationally induced particle production: Thermodynamics and
	kinetic theory.
	\newblock {\em Phys. Rev. D}, 90:043515, 2014.
	
\end{thebibliography}

\end{document}